\documentclass[useAMS,usenatbib]{mn2e}
\usepackage{times}
\usepackage{graphicx}

\newif\ifAMStwofonts
\AMStwofontstrue

\newcommand{\simlt}{\lower.5ex\hbox{$\; \buildrel < \over \sim \;$}}
\newcommand{\simgt}{\lower.5ex\hbox{$\; \buildrel > \over \sim \;$}}
\newcommand{\be}{\begin{equation}}
\newcommand{\ba}{\begin{eqnarray}}
\newcommand{\ee}{\end{equation}}
\newcommand{\ea}{\end{eqnarray}}

\title[Merging channel of massive galaxies since z$\sim$1]
{Constraints on the merging channel of massive galaxies since z$\sim$1}
\author[Ferreras et al.]  
{I. Ferreras$^1$\thanks{Email: i.ferreras@ucl.ac.uk}, 
I. Trujillo$^{2,3}$, E. M\'armol-Queralt\'o$^{2,3,4}$, 
P.~G. P\'erez-Gonz\'alez$^5$,\and
A. Cava$^{5,6}$, G. Barro$^7$, 
J. Cenarro$^8$, A. Hern\'an-Caballero$^9$, N. Cardiel$^5$, \and
J. Rodr\'\i guez-Zaur\'\i n$^{2,3}$, M. Cebri\'an$^{2,3}$\\
$^1$ Mullard Space Science Laboratory, University College London,
  Holmbury St Mary, Dorking, Surrey RH5 6NT, UK\\
$^2$ Instituto de Astrof\'\i sica de Canarias, C/ V\'\i a L\'actea
s/n, La Laguna, E-38200 La Laguna, Tenerife, Spain\\ 
$^3$ Departamento de Astrof\'\i sica,
Universidad de La Laguna, E-38205 La Laguna, Tenerife, Spain \\
$^4$ Institute for Astronomy, University of Edinburgh, Royal
Observatory, Blackford Hill, Edinburgh EH9 3HJ, UK\\
$^5$ Departamento de Astrof\'\i sica, Facultad de CC. F\'\i sicas, 
Universidad Complutense de Madrid, E-28040 Madrid, Spain\\
$^6$ Observatoire de Gen{\`e}ve, Universit{\'e} de Gen{\`e}ve, 51
Ch. des Maillettes, 1290 Versoix, Switzerland\\
$^7$ UCO/Lick Observatory, Department of Astronomy and Astrophysics,
University of California, Santa Cruz, CA 95064, USA\\
$^8$ Centro de Estudios de F\'\i sica del Cosmos de Arag\'on, 
Plaza San Juan 1, Planta 2, 44001 Teruel, Spain\\
$^9$ Instituto de F\'\i sica de Cantabria, CSIC-UC, 
Avenida de los Castros s/n, E-39005 Santander, Spain
}

\begin{document}
\date{Accepted 2014 July 15.  Received 2014 June 11; in original form 2013 December 18}
\pagerange{\pageref{firstpage}--\pageref{lastpage}} \pubyear{2014}
\maketitle
\label{firstpage}

\begin{abstract}
We probe the merging channel of massive galaxies over the z$=0.3-1.3$
redshift window by studying close pairs in a sample of $238$ galaxies
with stellar mass $\simgt 10^{11}$M$_\odot$, from the SHARDS survey.
SHARDS provides medium band photometry equivalent to low-resolution
optical spectra (R$\sim$50), allowing us to obtain extremely accurate
photometric redshifts (median $|\Delta z|/(1+z)\sim 0.55$\%) and to
improve the constraints on the age distribution of the stellar
populations. Our dataset is volume-limited, probing merger
progenitors with mass ratios 1:100 ($\mu\equiv M_{\rm sat}/M_{\rm cen}
=0.01$) out to z=1.3. A strong correlation is found between the age
difference of host and companion galaxy and stellar mass ratio, from
negligible age differences in major mergers to age differences
$\sim$4\,Gyr for 1:100 minor mergers. However, this correlation is
simply a reflection of the mass-age trend in the general
population. The dominant contributor to
the growth of massive galaxies corresponds to mass ratios $\mu\simgt
0.3$, followed by a decrease in the fractional mass growth rate
linearly proportional to $\log\mu$, at least down to $\mu\sim 0.01$,
suggesting a decreasing role of mergers involving low-mass companions,
especially if dynamical friction timescales are taken into account. A
simple model results in an upper limit for the average mass growth
rate of massive galaxies of $(\Delta M/M)/\Delta t\sim 0.08\pm
0.02$\,Gyr$^{-1}$, over the z$\simlt$1 range, with a $\sim 70$\%
fractional contribution from (major) mergers with $\mu\simgt 0.3$.
The majority of the stellar mass contributed by mergers does not
introduce significantly younger populations, in agreement with the
small radial age gradients observed in present-day early-type
galaxies.
\end{abstract}

%%%%%%%%%%%%%%%%%%%%%%%%%%%%%%%%%%%%%%%%%%%%%%%%
%%%%%%%%%%%%%%%%  Figure 1   %%%%%%%%%%%%%%%%%%%
%%%%%%%%%%%%%%%%%%%%%%%%%%%%%%%%%%%%%%%%%%%%%%%%
\begin{figure*}
  \includegraphics[width=140mm]{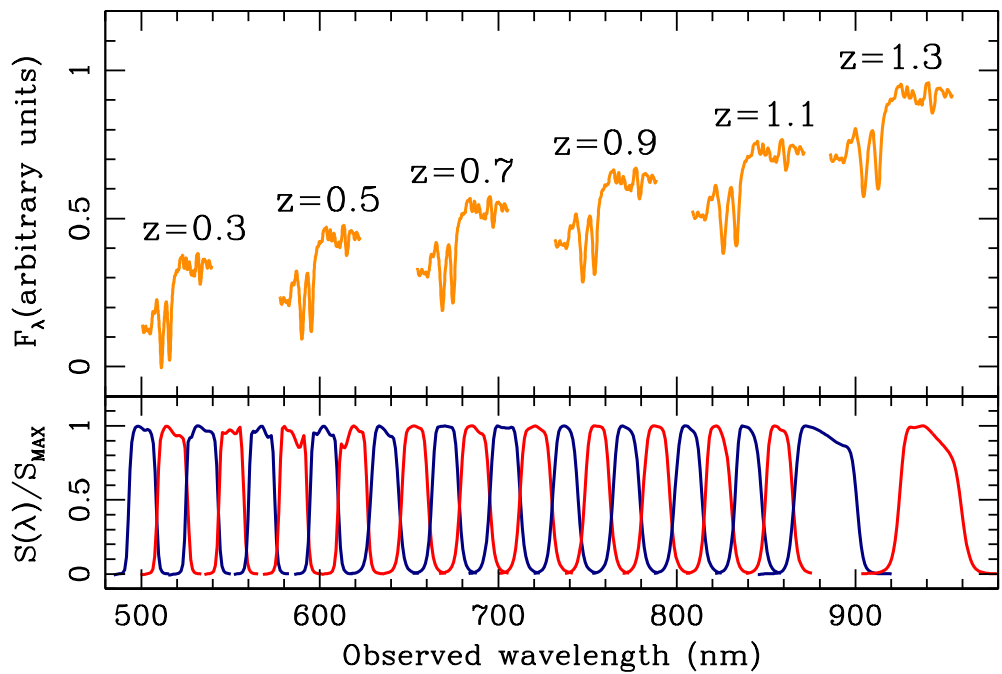}
  \caption{The spectral response of the SHARDS filters available for
    this study are shown (bottom, normalized in each case to the peak
    of the response). The upper panel shows the typical spectrum of an
    evolved stellar population around the age-sensitive
    4000\AA\ break, for a number of redshift values representative of
    our sample.}
  \label{fig:D4000}
\end{figure*}
%%%%%%%%%%%%%%%%%%%%%%%%%%%%%%%%%%%%%%%%%%%%%%%%

%%%%%%%%%%%%%%%%%%%%%%%%%%%%%%%%%%%%%%%%%%%%%%%%
\section{Introduction}

One of the possible scenarios for the formation and evolution of the
quiescent massive galaxies at low-redshift posits that after forming
the core of their structures in an early dissipative event at high
redshift (z$\simgt$2), these galaxies continue growing in mass and
size due to a continuous infall of material \citep[see,
  e.g.,][]{Naab:09}. In this picture, the main channel of growth is
provided by mergers with companion galaxies having a mass ratio 1:5
\citep{Oser:12}. The homogeneous age distribution of massive galaxies
out to z$\sim$1 \citep{ferr05,ferr09,PRS09} reveals that these mergers must
be relatively gas poor, without any significant star formation
\citep{ppg08,I3}. Numerical simulations suggest that the stellar
content of this accreted material should be mainly deposited in the
outer regions of the massive galaxy, with ages around 2.5\,Gyr older
than the component formed in-situ \citep{Lackner:12}. Observational
evidence of minor merger progenitors -- loosely defined as those where the
stellar mass ratio between host and companion is lower than 1:4 -- is
found in various systems, including stellar streams in the Milky
Way \citep[e.g.,][]{Bell:08}, and the presence of residual star formation in 
nearby early-type galaxies \citep[][]{Kav:07}.  Although several pieces of
observational evidence favour the above depicted scenario, the role of
minor mergers on the growth of massive galaxies is still poorly constrained.

In the z$\sim$0 Universe, the aftermath of merging activity can be
assessed via ultra-deep photometry and spectroscopy, probing the
radial gradients of the properties of stellar populations \citep[see,
  e.g.,][]{PSB:07}. However, the interpretation is hard because of the
challenging observations, the difficulty in circumventing the
age-metallicity degeneracy inherent in old stellar populations, and
the added complication of the presence of a halo of old, metal-poor
stars at galactocentric distances R$\simgt 4-8$R$_e$
\citep{FLB:12}. In fact, detailed surface brightness analyses of
nearby massive early-type galaxies reveal a complex structure, where
at least three components can be discerned: a compact core; an
intermediate region, and an outer halo \citep{Huang:13,Montes:14}. Separating
these components in nearby systems to derive the assembly history
poses a serious challenge.  Alternatively, it is possible to explore
at high redshift the stellar population properties of close pairs of
galaxies involving a massive galaxy (host) and a satellite (companion)
that will eventually collapse into the main body. This is the approach
followed in this paper. In contrast with studies where the selection
of merger progenitors is based on morphological features \citep[see,
  e.g.,][]{CC:06,Lotz:11,Bluck:12}, our selection is based on targeting pairs
of galaxies with a small projected separation {\sl and} located at the
same redshift, a more conservative approach \citep[see,
  e.g.][]{Patton:00,Rogers:09,Lin:10,LSJ:11}, however, requiring
accurate redshift estimates. In \citet{emqI,emqII}, a sample of
massive galaxies out to z$\simlt$2 was studied, finding a constant
fraction ($\sim$30\%) of massive galaxies with one, or more,
satellites and younger stellar populations in the satellites with
respect to the centrals. However, this sample was limited by the
uncertainties in the photometric redshift estimates based on broadband
photometry -- at the level of $|\Delta z|/(1+z)=5$\%. This paper
extends this work in two ways: Firstly by the use of improved
photometric data that allows for photometric redshifts that are an
order of magnitude more accurate.  Secondly, the data provides
low-resolution photo-spectra, resulting in better
constraints on  the age of the underlying stellar
populations of host and companion in merger progenitors.

The Survey for High-z Absorption Red and Dead Sources \citep[hereafter
  SHARDS, ][]{SHARDSI} comprises ultra-deep optical
spectro-photometric observations acquired at the 10.4m Gran Telescopio
Canarias (GTC) that provide an effective spectral resolution
$R=\lambda/\Delta\lambda\sim$50, with a set of 24 medium band
filters. This survey enables us to obtain accurate redshifts and to
probe the underlying stellar populations. We use here the SHARDS data
to define a sample of massive galaxies, with a robust characterization
of nearby companions as merger progenitors, allowing us to constrain
in detail the merging channel of massive galaxies since z$\sim$1. This
paper is structured as follows: In Section~2, we present the sample of
massive galaxies and the selection of host-companion pairs,
including a description of the methodology followed to obtain accurate
photometric redshifts, and the derivation of stellar ages. Section~3
compares the distribution of the ages of the underlying stellar
populations, followed by an estimate, in Sec.~4, of the contribution
in mass of these satellites to the growth of massive
galaxies. Finally, Sec.~5 gives an overview of our
conclusions. Throughout this paper, a standard $\Lambda$CDM cosmology
is adopted: $\Omega_m=0.3$, $\Omega_\Lambda=0.7$,
H$_0=70$\,km\,s$^{-1}$\,Mpc$^{-1}$ (or, equivalently, $h_{70}=1$).

%%%%%%%%%%%%%%%%%%%%%%%%%%%%%%%%%%%%%%%%%%%%%%%%
%%%%%%%%%%%%%%%%  Figure 2   %%%%%%%%%%%%%%%%%%%
%%%%%%%%%%%%%%%%%%%%%%%%%%%%%%%%%%%%%%%%%%%%%%%%
\begin{figure*}
  \includegraphics[width=150mm]{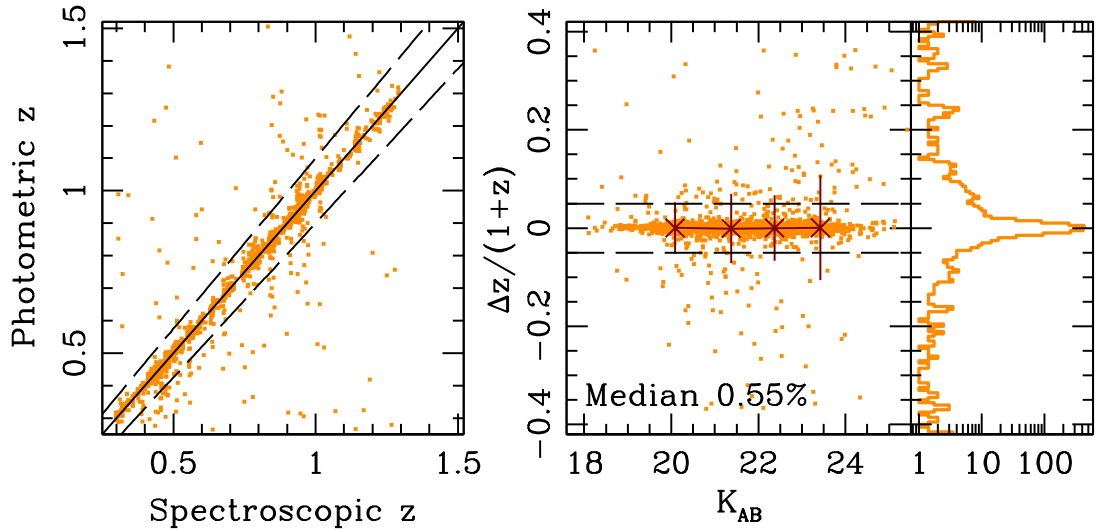}
  \caption{{\sl Left:} Comparison of photometric and spectroscopic
    redshifts for a subsample of 2792 galaxies with available
    spectroscopic data. The solid and dashed lines represent the 1:1
    correspondence, and an uncertainty of 5\%, respectively. {\sl
      Right:} The accuracy in the photometric redshifts is shown as a
    function of the total apparent $K$-band magnitude, including the 
    same 5\% interval as a dashed line (typical of standard photometric
    redshifts with broadband photometry alone). The rightmost panel
    shows the distribution of the photometric redshift accuracy in SHARDS.}
  \label{fig:photz}
\end{figure*}
%%%%%%%%%%%%%%%%%%%%%%%%%%%%%%%%%%%%%%%%%%%%%%%%

%%%%%%%%%%%%%%%%%%%%%%%%%%%%%%%%%%%%%%%%%%%%%%%%
\section{Sample Selection}

\subsection{Datasets}

SHARDS provides deep imaging (m$<$26.5\,AB at the $3\,\sigma$ level)
in a number of contiguous, medium band filters towards the GOODS-N
region, over a 130\,arcmin$^2$ field of view, split into two
pointings. The survey covers a 500--950\,nm spectral range in 24
medium-band filters (FWHM=15\,nm, except for the two reddest ones,
with FWHM=30\,nm). Tab.~\ref{tab:filters} gives an overview of the
SHARDS filters used and the depth and seeing of the observations
\citep[see][for details]{SHARDSI}.
The main science driver of the SHARDS survey is the
characterization of the stellar populations of massive galaxies at
z$>$1 by use of the Mg$_{\rm UV}$ spectral feature
\citep{SHARDSI,Daddi:05}. However, the data set also provides
information of the age-sensitive region around the 4000\,\AA\ break at
moderate redshift (see Fig.~\ref{fig:D4000}). In \citet{D4000} the
SHARDS data were used to analyze the D4000 index \citep[see,
  e.g.,][]{Balogh:99} at z$<$1.1. In this paper we concentrate on the
characterization of the close companions of massive galaxies using the
medium band photometry to secure accurate photometric redshifts and to
assess the age of the underlying stellar populations by way of
spectral fitting at a low spectral resolution. We find that the
optimal redshift range to extract stellar ages using the spectral
window around the 4000\AA\ break corresponds to
0.3$\simlt$z$\simlt$1.3, covering 5.3\,Gyr of cosmic time. Within the
adopted cosmology, this sample covers a total comoving volume of
$2.24\times 10^5h_{70}^{-3}$Mpc$^3$.

\begin{table}
\caption{Characteristics of the SHARDS filter set and observations.
  CWL is the central wavelength (in nm) of the filter for an
  angle of incidence AOI=10.5$^\circ$ (approximately that for the
  center of the FOV). The width of the filter is given as the FWHM of
  the response curve. m$_\mathrm{3\sigma}$ is the sensitivity limit,
  quoted  at the  3\,$\sigma$ level (AB mag). The average seeing is
  given in the last columns. P1 and P2 represent the individual pointings
  of the SHARDS observations.
}
\label{tab:filters}
\begin{center}
\begin{tabular}{lcccccc}
Filter & CWL & Width & \multicolumn{2}{c}{m$_\mathrm{3\sigma}$} & \multicolumn{2}{c}{seeing}\\
       & nm  & nm    & \multicolumn{2}{c}{AB}    &  \multicolumn{2}{c}{arcsec} \\
       &     &       & P1 & P2 & P1 & P2\\
\hline
F500W17  & 500.8 & 13.7 & 27.34 & 27.46 & 0.96 & 0.94\\
F517W17  & 519.7 & 14.6 & 27.16 & 27.42 & 0.79 & 0.93\\
F534W17  & 536.3 & 16.4 & 27.26 & 27.30 & 0.92 & 0.81\\
F551W17  & 552.1 & 12.3 & 27.15 & 27.19 & 0.83 & 0.78\\
F568W17  & 568.9 & 13.5 & 27.06 & 27.17 & 0.83 & 0.82\\
F585W17  & 586.5 & 14.0 & 27.15 & 27.24 & 0.97 & 0.89\\
F602W17  & 603.0 & 14.5 & 27.07 & 27.18 & 0.88 & 0.89\\
F619W17  & 618.9 & 14.8 & 27.06 & 27.18 & 0.85 & 0.90\\
F636W17  & 638.4 & 15.1 & 27.03 & 27.12 & 0.80 & 0.93\\
F653W17  & 653.1 & 14.6 & 27.09 & 27.15 & 0.98 & 1.00\\
F670W17  & 668.4 & 15.0 & 26.97 & 27.17 & 0.79 & 1.08\\
F687W17  & 688.2 & 15.1 & 26.98 & 27.04 & 0.83 & 0.90\\
F704W17  & 704.5 & 16.9 & 26.92 & 26.99 & 0.89 & 0.93\\
F721W17  & 720.2 & 18.0 & 26.88 & 26.94 & 0.94 & 1.02\\
F738W17  & 737.8 & 14.8 & 26.86 & 26.91 & 0.86 & 0.89\\
F755W17  & 754.5 & 14.6 & 26.87 & 26.90 & 0.93 & 0.94\\
F772W17  & 770.9 & 15.1 & 26.90 & 26.89 & 0.94 & 1.08\\
F789W17  & 789.0 & 15.3 & 26.79 & 26.84 & 0.96 & 0.93\\
F806W17  & 805.7 & 15.4 & 26.95 & 26.86 & 0.97 & 1.00\\
F823W17  & 825.4 & 14.5 & 26.88 & 26.81 & 0.82 & 0.96\\
F840W17  & 840.0 & 15.1 & 26.75 & 26.73 & 0.94 & 0.97\\
F857W17  & 856.4 & 15.4 & 26.88 & 26.77 & 0.73 & 0.98\\
F883W35  & 880.3 & 30.8 & 26.86 & 26.77 & 0.96 & 1.00\\
F941W33  & 941.0 & 30.0 & 26.63 & 26.64 & 0.91 & 0.95\\
\hline
\end{tabular}
\end{center}
\end{table}

In addition to the SHARDS data, we include the matched photometry from
the IRAC-selected ($[3.6\mu$m$]_{\rm AB}<24$) catalogue of
\citet{ppg08b} sources, for which UV to IR fluxes are available.  
Photometry in all UV-to-NIR bands was performed in elliptical
\citet{Kron:80} apertures. The best aperture was obtained for each galaxy
by averaging the values in the different bands, always imposing a minimum
semi-major axis equal to the worst seeing in our dataset (1.1\,arcsec).
Fluxes for the IRAC bands were measured within a 2\,arcsec radius circular 
aperture, applying an aperture correction for isolated sources. For sources
with companions closer than 2\,arcsec (roughly the FWHM of the IRAC PSF),
a deconvolution algorithm was applied, as explained in \cite{Barro:11}
The ground-based NIR photometry,
adopted from an update of the compilation of \citet{ppg08b}, is taken
from the deep observations of GOODS-N at Subaru/MOIRCS \citep{moircs}
and CFHT/WIRCam \citep{Wang:10}, for which a $5\sigma$ detection limit
of $K_{\rm AB}< 24.5$ is reached. In addition, we use the WFC3 NIR
photometry in the Y (F105W), J (F125W) and H (F160W) bands from the
CANDELS survey \citep[H$_{\rm AB}(5\,\sigma)<27$,][]{CANDELS,CANDELS1}.

%%%%%%%%%%%%%%%%%%%%%%%%%%%%%%%%%%%%%%%%%%%%%%%%
\subsection{SED Fitting: Photometric redshifts}
\label{ssec:Photz}

Photometric redshifts were estimated with the EAZY software
\citep{EAZY}, using the templates tweaked for emission-line galaxies
and medium-band photometry released in version 1.1.  We fitted the
available SEDs covering the wavelength range between the UV and the
MIR, including the SHARDS data. The medium band filters enable us to
obtain more accurate redshifts with respect to those based on
broadband photometry.  In \cite{SHARDSI}, we discussed in detail a
unique characteristic of our data: for observations using a given
filter, the passband varies along the FOV of the GTC/OSIRIS instrument
(resulting in a shift of the effective central wavelength, but keeping
the shape constant).  Hence, each galaxy in a SHARDS image ``sees'' a
different passband. In order to take into account the variation of the
central wavelength of the passband, a different execution of EAZY was
carried out for each individual galaxy, setting the filter
transmission files to the appropriate central wavelengths. This
procedure was essential to derive accurate photometric redshifts. In
fact, the estimates were a factor of 10 less precise when the shifts
were not taken into account. A more detailed description of the method
and results to estimate photometric redshifts will be presented in a
future paper (P\'erez-Gonz\'alez et al. 2014, in preparation). Given
that this paper focuses on the study of massive galaxies and their
companions at z$\simlt$1.3, we give photometric redshift quality
figures only within this redshift range.

%%%%%%%%%%%%%%%%%%%%%%%%%%%%%%%%%%%%%%%%%%%%%%%%
%%%%%%%%%%%%%%%%  Figure 3   %%%%%%%%%%%%%%%%%%%
%%%%%%%%%%%%%%%%%%%%%%%%%%%%%%%%%%%%%%%%%%%%%%%%
\begin{figure}
  \includegraphics[width=8.5cm]{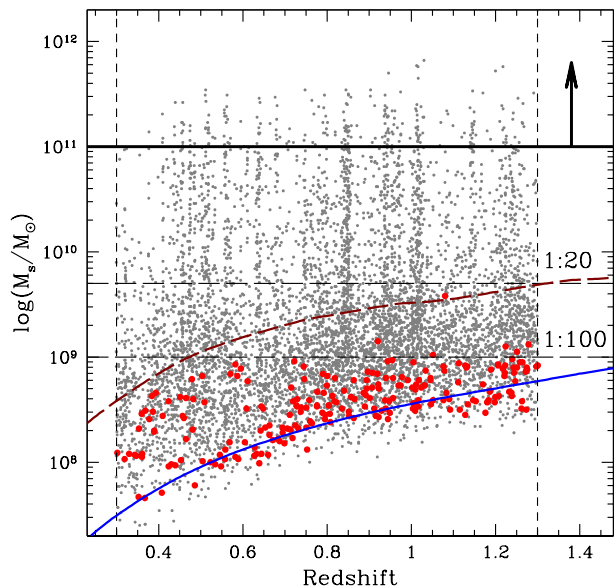}
  \caption{Stellar mass vs redshift diagram of the SHARDS sources
    used in this paper. A galaxy is classified as massive if the
    stellar mass is M$_s>10^{11}$\,M$_\odot$ (arrow and thick
    horizontal line). The horizontal dashed lines show the limit for
    1:20 and 1:100 merger progenitors. The blue solid line shows a
    $K_{\rm AB}=25$ simple stellar population with age 100\,Myr,
    whereas the red dashed line corresponds to a $K_{\rm AB}=25$ population formed at
    $z_{\rm FOR}=3$ (both at solar metallicity, Chabrier IMF, using
    the models of Bruzual \& Charlot 2003). The red dots correspond to
    sources with K$_{\rm AB}\sim$25.}
  \label{fig:Klim}
\end{figure}
%%%%%%%%%%%%%%%%%%%%%%%%%%%%%%%%%%%%%%%%%%%%%%%%

Fig.~\ref{fig:photz} shows the accuracy of our photometric redshift
catalogue, for a subsample of 2,792 galaxies with available
spectroscopic redshifts from the literature \citep[down to $I\sim$25,
  see, e.g.,][]{Cowie:04,Wirth:04,Cooper:12}. Over the 0.3$<$z$<$1.3
range covered in this paper, accurate photometric redshifts are
obtained, largely due to the presence of the prominent
4000\,\AA\ break within the spectral coverage of the medium-band
filter (see Fig.~\ref{fig:D4000}). The left-hand panel of
Fig.~\ref{fig:photz} compares spectroscopic versus photometric
redshift, with the solid line tracking a 1:1 correspondence, and the
dashed lines spanning a 5\% uncertainty, typical of standard
photometric redshifts involving broadband data
\citep[e.g.][]{ppg08b}. The right-hand panel shows the photo-z
accuracy with respect to total $K$-band magnitude, and the histogram
with the narrow distribution. Note we use a logarithmic axis since the
accuracy results in a highly peaked histogram on a linear scale. The
dots represent individual galaxies, and the red crosses are binned
estimates (at a fixed number of galaxies per bin), including the RMS
within each bin as an error bar. The median of the distribution of
$|\Delta z|/(1+z)$ is $0.55$\%, with a $1.6$\% fraction of
catastrophic failures, using a very conservative definition, as those
redshift estimates with an accuracy outside of $3\times$ the RMS of
the distribution.

%%%%%%%%%%%%%%%%%%%%%%%%%%%%%%%%%%%%%%%%%%%%%%%%
\subsection{SED Fitting: Stellar population properties}
\label{ssec:StPops}

We model the SED of the parent sample using stellar population
synthesis models to determine their stellar properties, primarily
stellar masses and ages.  These are critical for the selection of the
main sample of massive galaxies and to analyze the age
differences between them and their close companions.  Briefly, the
method involves a comparison of the SHARDS medium-band photometry with
a large volume of composite models derived from the stellar population
synthesis library of \citet{BC03} -- a \citet{Chab:03} stellar initial
mass function (IMF) is assumed throughout this paper. Although recent
claims suggest a more bottom-heavy IMF in massive galaxies \citep[see,
  e.g.,][]{Cen:03,VdKC:10,Ferr:13}, its effect on the stellar M/L, and
therefore on the stellar masses derived from the photometry, may not
be as important, except for systems with very high velocity dispersion
\citep{FLB:13, Laesker:13}.

%%%%%%%%%%%%%%%%%%%%%%%%%%%%%%%%%%%%%%
%%%%%%%%%%   TABLE 1   %%%%%%%%%%%%%%%
%%%%%%%%%%%%%%%%%%%%%%%%%%%%%%%%%%%%%%
\begin{table}
\caption{Parameters defining the model grids for the analysis of
  stellar ages, based on \citet{BC03} models, and dust extinction from
  \citet{CCM89}.}
\label{tab:pars}
\begin{center}
\begin{tabular}{lccr}
\hline
%& & \multicolumn{3}{c}{Exposure Time (s)}\\
Parameter & Symbol & Range & N\\
\hline
\multicolumn{4}{c}{SIMPLE STELLAR POPULATIONS (SSP)}\\
Metallicity & $\log Z/Z_\odot$& $[-2.0,+0.3]$ & 6\\
Age         & $\log t/{\rm Gyr}$ & $[-2,\log t_U/{\rm Gyr}]$ & 64\\
Emission Lines & EW([O\,{\textsc II}]) & $[0,30]$\,\AA & 12\\
Dust Reddening & E(B--V) & $[0,0.5]$ & 12\\
\hline
\multicolumn{4}{c}{$\tau$ MODELS ($\tau$SF)}\\
Metallicity & $\log Z/Z_\odot$& $[-2.0,+0.3]$ & 6\\
SF timescale & $\log \tau_{\rm SF}/{\rm Gyr}$ & $[-1,1]$ & 16\\
Formation epoch & $t_{\rm FOR}/t_U$ & [0,1) & 16\\
Emission Lines & EW([O\,{\textsc II}]) & $[0,30]$\,\AA & 12\\
Dust Reddening & E(B--V) & $[0,0.5]$ & 12\\
\hline
\multicolumn{4}{c}{CONSTANT SF MODELS (CST)}\\
Metallicity & $\log Z/Z_\odot$& $[-2.0,+0.3]$ & 6\\
Truncation timescale & $\log \tau_t/{\rm Gyr}$ & $[-1,1]$ & 16\\
Formation epoch & $t_{\rm FOR}/t_U$ & [0,1) & 16\\
Emission Lines & EW([O\,{\textsc II}]) & $[0,30]$\,\AA & 12\\
Dust Reddening & E(B--V) & $[0,0.5]$ & 12\\
\hline
\multicolumn{3}{r}{Total Models} & 497,664\\
\end{tabular}
\end{center}
$t_U$ is the age of the Universe at the redshift of the
galaxy under study.
\end{table}
%%%%%%%%%%%%%%%%%%%%%%%%%%%%%%%%%%%%%%

%%%%%%%%%%%%%%%%%%%%%%%%%%%%%%%%%%%%%%%%%%%%%%%%
%%%%%%%%%%%%%%%%  Figure 4   %%%%%%%%%%%%%%%%%%%
%%%%%%%%%%%%%%%%%%%%%%%%%%%%%%%%%%%%%%%%%%%%%%%%
\begin{figure}
  \includegraphics[width=8.5cm]{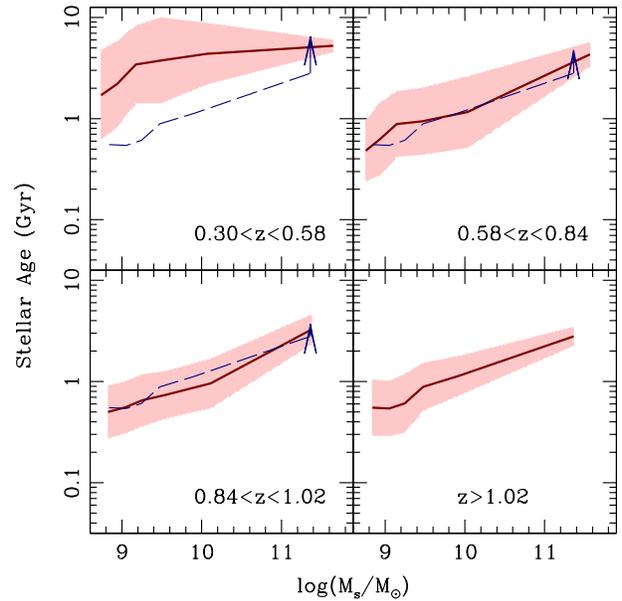}
  \caption{Mass-age relationship split into quartiles
  in redshift. The red lines and shading give the median 
  and RMS scatter, respectively. The dashed lines give, for
  reference, the mass-age relation in the highest redshift bin.
  The arrows at the massive end correspond to the cosmic time
  spanned between the different redshift bins. 
}
  \label{fig:MvsAge}
\end{figure}
%%%%%%%%%%%%%%%%%%%%%%%%%%%%%%%%%%%%%%%%%%%%%%%%

%%%%%%%%%%%%%%%%%%%%%%%%%%%%%%%%%%%%%%%%%%%%%%%%
%%%%%%%%%%%%%%%%  Figure 5   %%%%%%%%%%%%%%%%%%%
%%%%%%%%%%%%%%%%%%%%%%%%%%%%%%%%%%%%%%%%%%%%%%%%
\begin{figure*}
  \includegraphics[width=140mm]{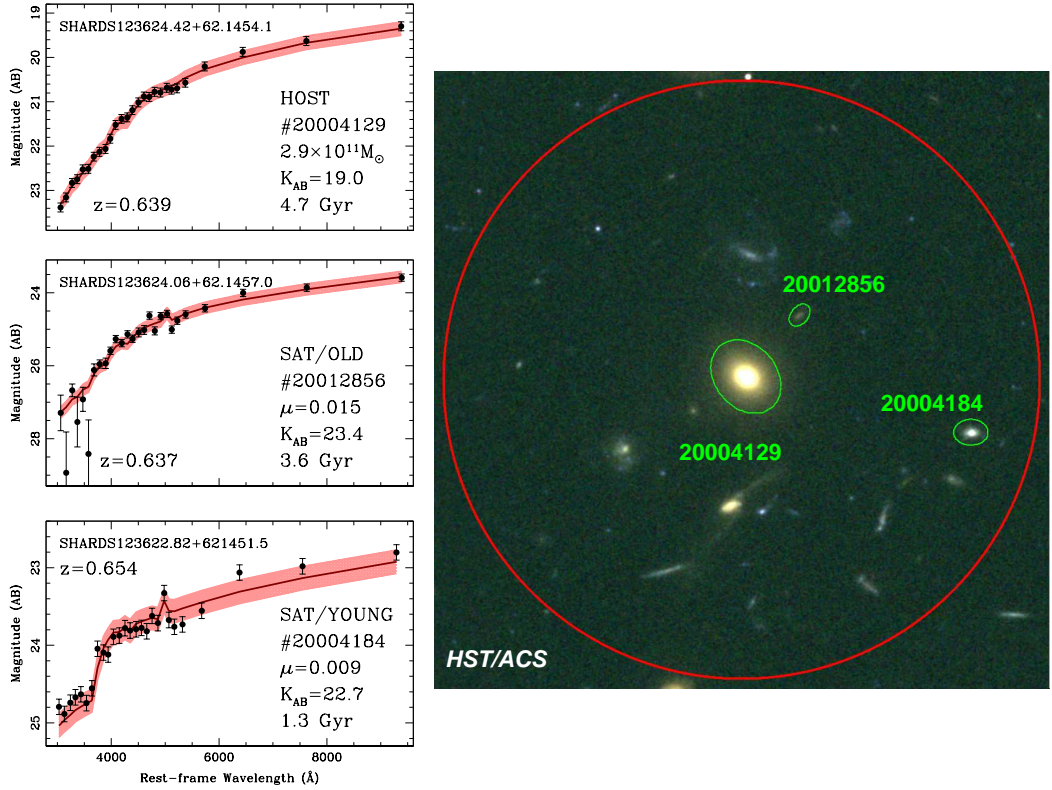}
  \caption{Example of a system with SEDs and best fit. The host is
    massive galaxy SHARDS123624.42+621454 (z=0.639, M$_s=2.9\times
    10^{11}$M$_\odot$). The $30\times 30$\,arcsec$^2$ image ({\sl
      right}) is an RGB colour composite from the GOODS-N {\sl
      HST/ACS} archival data through the F606W, F775W, and F850LP
    passbands. The red circumference extends over the
    $100h_{70}^{-1}$\,kpc search radius.  On the left, the observed
    data points (black circles) are compared with the best fit models
    (red lines). Each galaxy is labelled by its SHARDS ID,
    $K_{\rm AB}$ apparent magnitude, best-fit stellar age, and either
    stellar mass (host) or mass ratio (companion).}
  \label{fig:example}
\end{figure*}
%%%%%%%%%%%%%%%%%%%%%%%%%%%%%%%%%%%%%%%%%%%%%%%%

Three sets of model grids are considered, comprising simple stellar
populations (i.e. a single burst population); an exponentially
decaying star formation history ($\tau$ model), or a constant star
formation truncated after a timescale chosen as a free parameter. In
addition to the continuum from the stellar population synthesis
models, we include emission lines corresponding to a H{\textsc II}
region \citep{Aller:84}. The weighting of the emission line spectrum
is parameterised by the equivalent width of the [O\,{\textsc II}]
line. We note that the nebular continuum is neglected here, since we
are dealing with a general sample of galaxies, where
luminosity-weighted ages are always older than 0.1\,Gyr. The
contribution from the nebular continuum at those ages is expected to
be negligible \citep{Anders:03}. The models are also subject to a
Milky Way-type extinction, following the prescription of
\citet{CCM89}. The range of parameters used in the definition of the
templates are summarised in Tab.~\ref{tab:pars}. The library of nearly
half a million templates is integrated with the response of all the
available filters, including a number of wavelength offsets in steps
of 50\AA\ (smaller changes in central wavelength cause variations at a
much lower level than the photometric uncertainties). These offsets
are needed to take into account the variation of the central
wavelength of the filter with respect to the position of the source on
the field of view \citep[see][for details]{SHARDSI}.  For each galaxy,
all models are run for the specific redshift and SHARDS filter offsets
(which are unique for each galaxy and depend of the position of the
image with respect to the optical axis), and compared with the
observed fluxes measured in the SHARDS data. The comparison is used
to derive a standard $\chi^2$ statistic.  A probability distribution
function for the parameters is extracted from the $\chi^2$-derived
likelihood -- i.e. ${\cal P}\propto\exp(-\frac{1}{2}\Delta\chi^2)$.
For the type of sampling explored in this paper -- where the number of
gridpoints has to be constrained -- probability-weighted values give
more robust estimations than the actual gridpoint where the highest
value of the likelihood is reached.  Hereafter, all best-fit
parameters are quoted as probability-weighted quantities.  We
emphasize that although this method is fairly rigid in terms of the
star formation histories allowed, it is much more robust than spectral
fitting methods based on partial searches of parameter space, such as
the Monte Carlo Markov Chain approach \citep[see,
  e.g.,][]{Starlight}. We performed 500 simulations of arbitrary star
formation histories to check that our retrieved ages were consistent
within the error bars with the input values.  Hereafter, stellar age
is weighted with respect to the star formation rate, $\psi(t)$,
namely:
\begin{equation}
{\rm Age}\equiv\langle{\rm Age}\rangle_\psi = \frac{\int [t_U(z)-t]\psi(t)dt}{\int \psi(t)dt},
\end{equation}
where the integrals extend between the formation time, $t_{\rm FOR}$,
and the age of the Universe at the redshift of the galaxy, $t_U(z)$.
We note that the age difference between SSP, $\tau$ and
constant SF models is rather small: the difference $\Delta \equiv \log
({\rm Age}_{\rm CST})-\log(\rm {Age}_{\rm SSP})$ is $\Delta=0.14\pm
0.84$ ($1\sigma$) for all galaxies, and $\Delta=0.07\pm 0.43$ for
massive galaxies. In the Appendix we elaborate on the comparison between
the age estimates according to these models, confirming that the
trends found do not depend on the parameterisation of the star formation
histories. Hereafter, the quoted ages and stellar masses correspond
to those derived from the $\tau$ models.

%%%%%%%%%%%%%%%%%%%%%%%%%%%%%%%%%%%%%%%%%%%%%%%%
%%%%%%%%%%%%%%%%  Figure 6  %%%%%%%%%%%%%%%%%%%
%%%%%%%%%%%%%%%%%%%%%%%%%%%%%%%%%%%%%%%%%%%%%%%%
\begin{figure}
  \includegraphics[width=8.5cm]{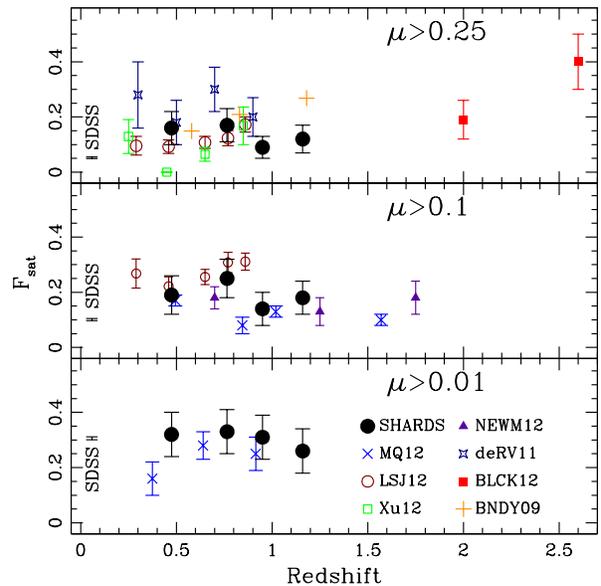}
  \caption{Redshift evolution of the fraction of close companions of
    massive galaxies ($M_s> 10^{11}M_\odot$), for three choices of the
    stellar mass ratio between satellite and central ($\mu$).  Our
    results (solid circles, SHARDS) are compared with recent results
    from the literature: \citet['x' signs, MQ12]{emqI}; Ruiz et al. (in preparation,
    estimates at z=0, labelled 'SDSS'); \citet[open circles, LSJ12]{LSJ:12};
    \citet[open squares, Xu12]{Xu:12}; \citet[triangles, NEWM12]{Newman:12};
    \citet[stars, deRV11]{deRavel:11}; \citet[filled squares, BLCK12]{Bluck:12};
    \citet['+' signs, BNDY09]{Bundy:09}.}
  \label{fig:fsat}
\end{figure}
%%%%%%%%%%%%%%%%%%%%%%%%%%%%%%%%%%%%%%%%%%%%%%%%

%%%%%%%%%%%%%%%%%%%%%%%%%%%%%%%%%%%%%%
%%%%%%%%%%   TABLE 2   %%%%%%%%%%%%%%%
%%%%%%%%%%%%%%%%%%%%%%%%%%%%%%%%%%%%%%
\begin{table*}
\caption{Fraction of massive galaxies with close companions at various
  redshifts.  For each redshift bin, $N_{\rm cen}$ is the total number
  of massive (central) galaxies; $F_{\rm obs}$ is the observed
  fraction of those with nearby sources. A simulation to correct for
  chance alignments due to the uncertainty in the estimates of
  redshift is given by $S_{\rm simul}$; with the corrected fraction
  being $F_{\rm sat}$. A final correction, $S_{\rm clust}$ has to be
  made to account for galaxy clustering.  The final, corrected
  fraction of satellites is given by the last column, $F_{\rm
    sat,clust}$ (see text for details). }
\label{tab:sims}
\begin{center}
\begin{tabular}{lcccccc}
\hline
Redshift range & $N_{\rm cen}$ & $F_{\rm obs}$ & $S_{\rm simul}$ & $F_{\rm sat}$ & 
$S_{\rm clust}$ & $F_{\rm sat,clust}$\\
\hline 
\multicolumn{5}{l}{$\mu\equiv M_{\rm sat}/M_{\rm cen}\geq 0.25$}\\
$0.30 - 0.65$ & $59$ & $0.19\pm 0.05$ & $0.03\pm 0.02$ & $0.16\pm 0.05$ & $0.03\pm 0.03$ & $0.16\pm 0.06$\\
$0.65 - 0.88$ & $62$ & $0.21\pm 0.05$ & $0.04\pm 0.02$ & $0.18\pm 0.05$ & $0.05\pm 0.03$ & $0.17\pm 0.06$\\
$0.88 - 1.02$ & $61$ & $0.12\pm 0.04$ & $0.04\pm 0.02$ & $0.08\pm 0.04$ & $0.03\pm 0.02$ & $0.09\pm 0.04$\\
$1.02 - 1.30$ & $56$ & $0.13\pm 0.05$ & $0.02\pm 0.02$ & $0.11\pm 0.05$ & $0.01\pm 0.02$ & $0.12\pm 0.05$\\
\hline
\multicolumn{5}{l}{$\mu\equiv M_{\rm sat}/M_{\rm cen}\geq 0.1$}\\
$0.30 - 0.65$ & $59$ & $0.24\pm 0.06$ & $0.06\pm 0.03$ & $0.19\pm 0.07$ & $0.06\pm 0.03$ & $0.19\pm 0.07$\\
$0.65 - 0.88$ & $62$ & $0.32\pm 0.06$ & $0.07\pm 0.03$ & $0.27\pm 0.07$ & $0.09\pm 0.04$ & $0.25\pm 0.07$\\
$0.88 - 1.02$ & $61$ & $0.21\pm 0.05$ & $0.07\pm 0.03$ & $0.15\pm 0.06$ & $0.08\pm 0.04$ & $0.14\pm 0.06$\\
$1.02 - 1.30$ & $56$ & $0.20\pm 0.05$ & $0.04\pm 0.03$ & $0.17\pm 0.06$ & $0.03\pm 0.03$ & $0.18\pm 0.06$\\
\hline
\multicolumn{5}{l}{$\mu\geq 0.01$}\\
$0.30 - 0.65$ & $59$ & $0.44\pm 0.06$ & $0.17\pm 0.05$ & $0.33\pm 0.08$ & $0.18\pm 0.05$ & $0.32\pm 0.08$\\
$0.65 - 0.88$ & $62$ & $0.48\pm 0.06$ & $0.20\pm 0.05$ & $0.35\pm 0.08$ & $0.22\pm 0.05$ & $0.33\pm 0.08$\\
$0.88 - 1.02$ & $61$ & $0.48\pm 0.06$ & $0.21\pm 0.05$ & $0.34\pm 0.08$ & $0.25\pm 0.05$ & $0.31\pm 0.08$\\
$1.02 - 1.30$ & $56$ & $0.36\pm 0.06$ & $0.15\pm 0.05$ & $0.25\pm 0.08$ & $0.14\pm 0.05$ & $0.26\pm 0.08$\\
\hline
\end{tabular}
\end{center}
\end{table*}
%%%%%%%%%%%%%%%%%%%%%%%%%%%%%%%%%%%%%%

%%%%%%%%%%%%%%%%%%%%%%%%%%%%%%%%%%%%%%
\subsection{Sample Selection}

Based on the best-fit stellar masses obtained from the modelling of
the stellar populations, we define our sample of (host) massive
galaxies as those systems with stellar mass $M_s\geq 10^{11}M_\odot$.
To determine the range of mass ratios between host and companion, we
estimate the completeness level of the parent catalogue as a function
of stellar mass.  Fig.~\ref{fig:Klim} shows the distribution of
stellar mass over the redshift range under consideration (vertical
dashed lines). To assess the completeness level, we compare our sample
with synthetic models that have an apparent magnitude $K_{\rm AB}=25$
(i.e. close to the limit of our sample). The dashed red line traces an old
population formed at z$_{\rm FOR}=3$, whereas the blue line
corresponds to a 100\,Myr age.  We infer a very conservative
completeness limit at a mass ratio of 1:20 if a wide range of stellar
ages are considered.  Due to the strong correlation between age and
stellar mass found in galaxies over a wide redshift range \citep[see,
  e.g.,][]{ppg08}, we also propose a more realistic limit of
1:100. The red dots are real data at the chosen limit of $K_{\rm
  AB}=25$, illustrating the fact that most of these galaxies have
young stellar ages.

Given the strong colour (i.e. star formation rate/age) bimodality as a
function of stellar mass \citep[i.e. red sequence vs. blue cloud, see,
  e.g.,][] {Bell:03,Baldry:04,Faber:07,Arnouts:07,Brammer:11}, we
expect the low-mass companions to be young (blue) star-forming
systems.  To illustrate and quantify this point, Fig.~\ref{fig:MvsAge}
plots the distribution of stellar age against mass for the whole
sample. The data are binned in four redshift intervals. The lines and
shaded regions trace the median and RMS scatter, respectively. Note
the strong age-mass correlation, and its evolution with redshift. The
dashed lines overlay the mass-age relation from the highest redshift
bin, with the vertical arrows extending over the difference in cosmic
time between redshift bins. These arrows confirm that passive
evolution is the main driver of the photometric properties of massive
galaxies at z$\simlt$1. Fig.~\ref{fig:MvsAge} shows that for systems
that would correspond to merger progenitors with a mass ratio below
1:20 (i.e. concerning companions at the low-mass end of the sample),
we should not expect {\sl old} ($\simgt 3-4$\,Gyr) populations in the
companions. In fact, for redshifts above the median value of the
sample (z$>$0.85), the fraction of $M_s<5\times 10^9M_\odot$ galaxies
with ages older than 3\,Gyr (2\,Gyr) is just 0.14\% (0.90\%). In
Sec.~\ref{sec:ages} we explore the completeness of the sample in more
detail.

We note that the present analysis is mainly geared towards the
determination of stellar ages. The models give significantly more
accurate error bars in age than metallicity: only six
independent metallicities are used in the modelling, and the age
range available, given the redshift distribution of the sample, allows
for more accurate age constraints than with low-redshift systems. A
typical example, SHARDS123624.42+621454.1 (z=0.639) is presented in
Fig.~\ref{fig:example}. The colour image is a $30^{\prime\prime}\times
30^{\prime\prime}$ RGB composite combining the F606W, F775W and
F850LP passbands from the {\sl HST}/ACS GOODS-N 
images\footnote{\tt http://archive.stsci.edu/prepds/goods/}. The red
circle extends over a projected radius of $100h_{70}^{-1}$\,kpc. The
central system is a $2.9\times 10^{11}$M$_\odot$ spheroidal galaxy,
surrounded by two close companions. The low-resolution SHARDS spectral
energy distribution of host and companions is shown on the
left, along with (1\,$\sigma$) error bars. The model best fit is given
for each galaxy as a red shading. The best fit ages are also included in
each diagram.

%%%%%%%%%%%%%%%%%%%%%%%%%%%%%%%%%%%%%%%%%%%%%%%%
\subsection{Final Clean-up}

The preliminary set of $285$ massive galaxies is visually inspected,
using the publicly available {\sl HST}/ACS images.  The objects are
classified as ``unresolved'' (i.e. PSF-like); spheroidal (including
both E and S0 morphologies); disc; irregular/merger and artefacts in
the selection (usually spurious detections next to a bright star).
The artefacts were removed from the catalogue, and the
borderline-unresolved sources were inspected in more detail, removing
those that are stars. By comparing the best fits from stellar
population synthesis (see sect.~\ref{ssec:StPops}) with a QSO template
\citep{vdBerk:01} or a set of stellar spectra \citep{Pickles:98}; we
found that all ``unresolved'' objects brighter than $K_{\rm AB}<17.5$
($8$ found in our sample) were better fit by a QSO spectrum, and,
therefore, removed from the final catalogue. The final set comprises
$238$ massive galaxies: $48$\% are spheroidal/unresolved; $45$\% are
disk-like; $7$\% have a merger/irregular morphology. The sample of
massive galaxies has $224$ systems with an available spectroscopic
redshift, allowing us to asses for this subsample an accuracy of
$|\Delta z|/(1+z)=0.46$\% (median), when using photometric redshifts.

%%%%%%%%%%%%%%%%%%%%%%%%%%%%%%%%%%%%%%%%%%%%%%%%
%%%%%%%%%%%%%%%%  Figure 6   %%%%%%%%%%%%%%%%%%%
%%%%%%%%%%%%%%%%%%%%%%%%%%%%%%%%%%%%%%%%%%%%%%%%
\begin{figure*}
  \includegraphics[width=8cm]{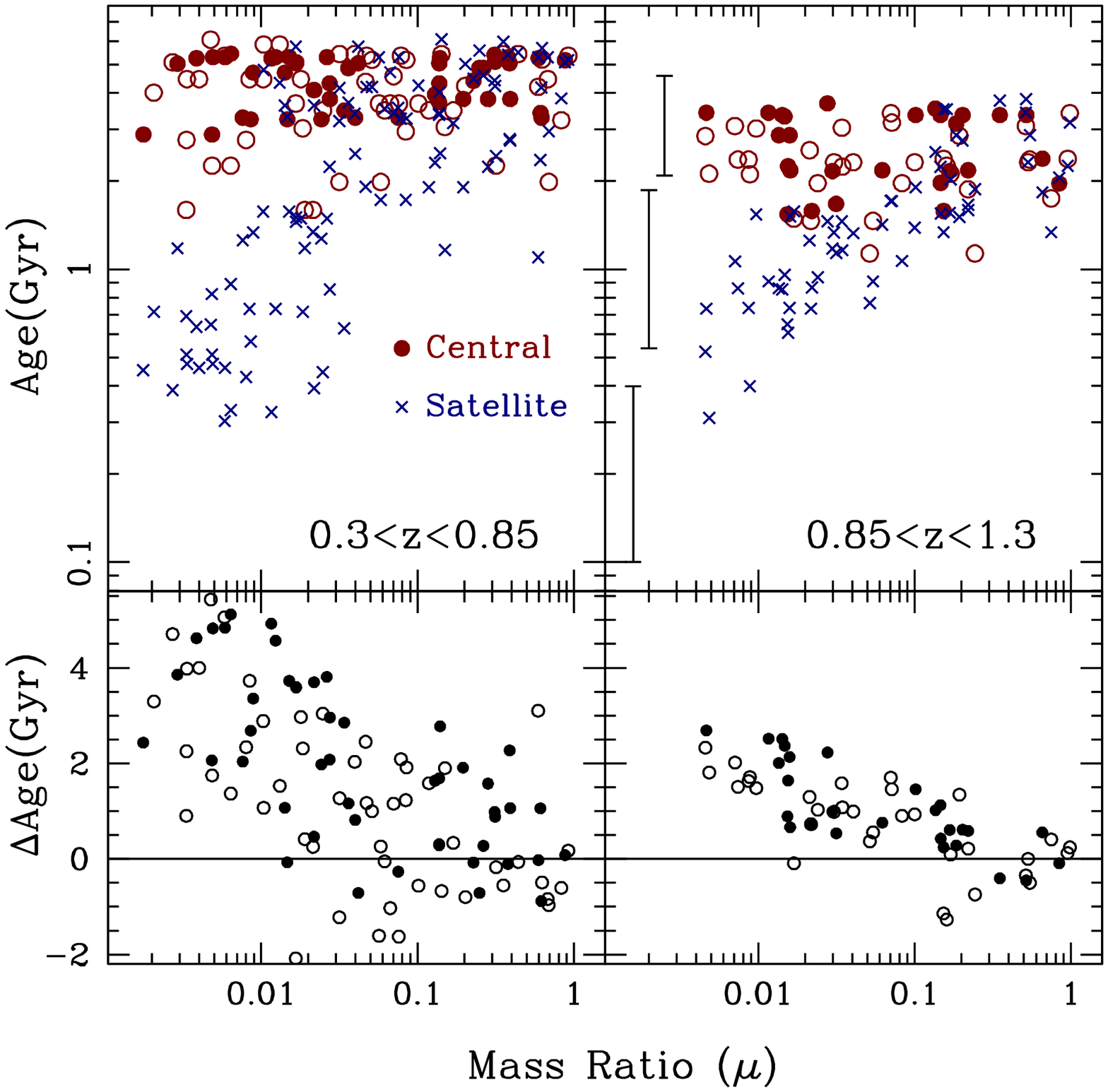}
  \includegraphics[width=8cm]{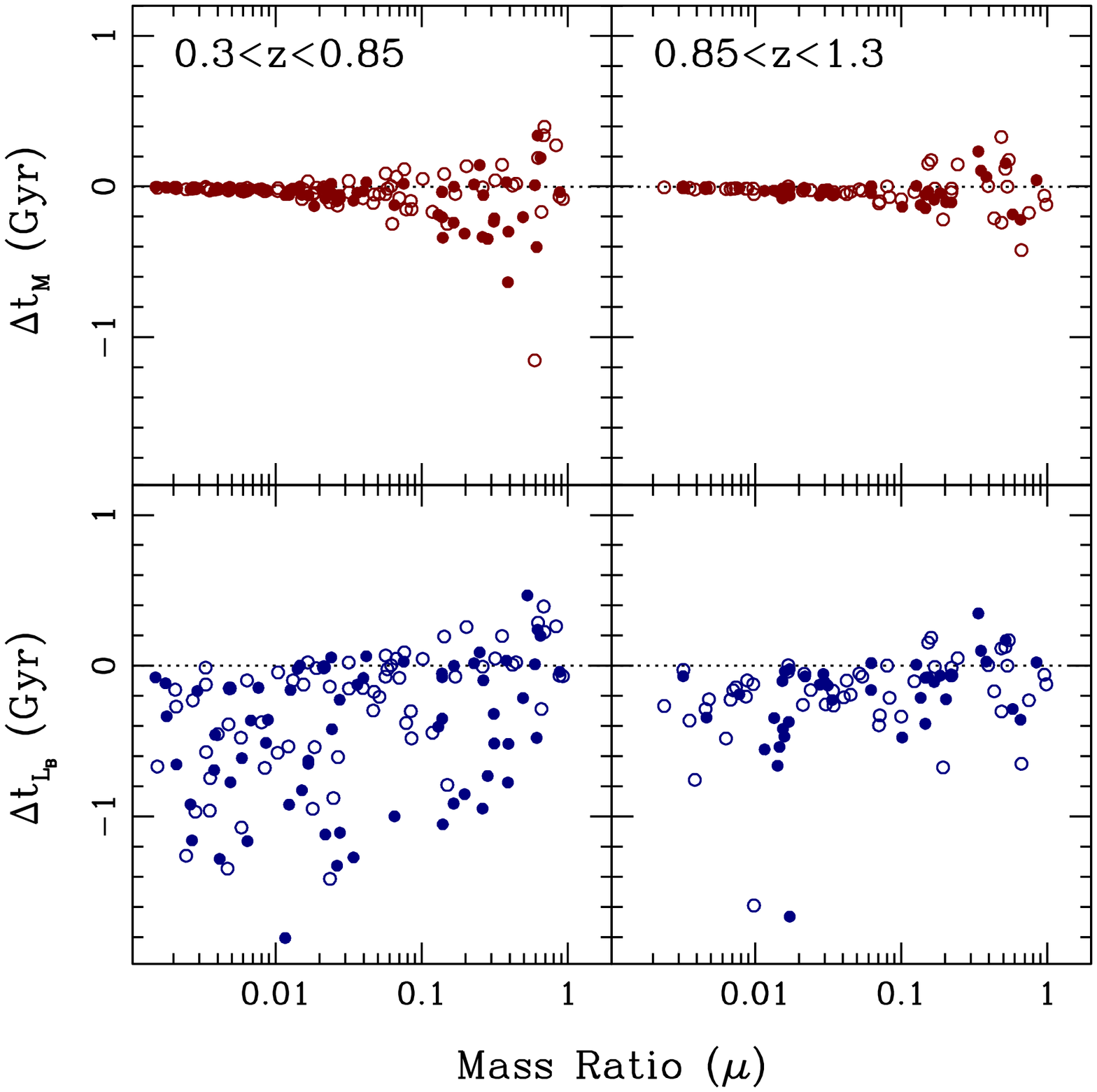}
  \caption{{\sl Left:} Age difference between host (red circles) and
    companion galaxies (blue crosses) for the exponentially decaying
    ($\tau$) models. The centrals are split with respect to visual
    morphology (solid circles for E/S0; open circles for
    disc/irregular).  The top panels show the individual ages, the
    bottom panels show the age difference between them. The sample is
    split at the median value of the redshift distribution. 
    {\sl Right:} Rejuvenation of the central galaxy assuming the close
    companion eventually merges with it. The age difference, $\Delta
    t\equiv$Age(After Merger)$-$Age(Before Merger), is weighted with
    respect to the stellar mass content (top) or the rest-frame
    $B$-band luminosity (bottom). The sample is split with respect to
    redshift at the median of the distribution.  }
  \label{fig:ages}
\end{figure*}
%%%%%%%%%%%%%%%%%%%%%%%%%%%%%%%%%%%%%%%%%%%%%%%%

%%%%%%%%%%%%%%%%%%%%%%%%%%%%%%%%%%%%%%%%%%%%%%%%
\section{Selection of Host-Companion Pairs}
\label{sec:Pairs}

For each of the $238$ massive galaxies, we define  merger
progenitors as those with a nearby satellite. In this paper,
a galaxy is considered a close companion if it is located at a
projected distance $\Delta r_\perp\leq 100h_{70}^{-1}$\,kpc, with a
redshift difference $|\Delta z|\leq 0.011\times (1+z)$, i.e. within the
2\,$\sigma$ uncertainty of the photometric redshift estimates.  
We decided to extend the interval in redshift to 2\,$\sigma$
because of the high accuracy of the photometric redshifts,
in constrast with previous works.
We note that 38\% of the companions have a spectroscopic redshift,
from which we derive a photometric redshift
accuracy of $|\Delta z|/(1+z)=0.75$\% (median).
In order to assess
the contamination from foreground/background sources because of the
use of photometric redshifts, we follow the methodology laid out in
\citet{emqI} (see their \S\S3.1). In a nutshell, a Monte Carlo method
is applied, where mock catalogues are generated by randomly placing
massive galaxies within the survey volume, keeping the same redshift
distribution as the original catalogue. The fraction of companions is
computed for each mock sample. We find that the Monte Carlo runs
give convergent fractions within the error bars after 1000 realizations.
The fraction of
simulated galaxy pairs is defined as $S_{\rm simul}$. We can use this
distribution to correct for the effect of contaminants in the true
sample \citep[see][for details]{emqI}.  If $F_{\rm obs}$ is the
observed fraction of close companions, the corrected one ($F_{\rm sat}$) is:
\begin{equation}
F_{\rm sat} = \frac{F_{\rm obs}-S_{\rm simul}}{1-S_{\rm simul}}
\label{eq:Fsat}
\end{equation}
Tab.~\ref{tab:sims} shows the fraction of massive ($\log
M_s/M_\odot>11$) galaxies with nearby satellites, for three different
choices of the stellar mass ratio between companion and host,
$\mu\equiv M_{\rm comp}/M_{\rm host}$.  We include in the table the
results corresponding to the Monte Carlo simulations to account for
spurious pairs.  Note that this correction stays at the level of
$\sim$5\% for pairs with $\mu\geq 0.1$ and $\sim$20\% for
$\mu\geq 0.01$. The 68\% ($1\sigma$ equivalent) error bars are derived from
a binomial distribution, following the favoured beta distribution
generator presented in \citet{Cameron:11}. In addition, we need to
take into account the effect of clustering. We perform a similar
estimate of the fraction of companions, as described above, for a set
of annuli from $100h_{70}^{-1}$\,kpc -- which is the fiducial maximum
distance chosen in the paper for the selection of pairs -- out to
$500h_{70}^{-1}$\,kpc, at fixed solid angle per annulus, equivalent to
the search area. Measurements in the field would yield no significant
variation in the number of galaxy pairs over this range, whereas a
highly clustered environment would display a monotonic decrease of the
number of satellites with increasing radius. We define the
environment-corrected value of $F_{\rm sat}$ as the one corresponding
to $500h_{70}^{-1}$\,kpc, including an error bar determined by the
RMS scatter of the satellite fraction between $400h_{70}^{-1}$ and
$500h_{70}^{-1}$\,kpc. These results are quoted under the column named
$S_{\rm clust}$ in Tab~\ref{tab:sims}. The last column ($F_{\rm
  sat,clust}$) gives the final result of our environment-corrected
fraction of satellites, taking into account the effect of
clustering. Note the increase in the clustering fraction around z$\sim
0.8-0.9$, where a structure is evident in the redshift distribution of
the sources towards the GOODS-N region. The final fraction of
satellites is found not to depend on redshift within the range
covered, with  $F_{\rm sat,corr}=0.19\pm 0.08$ (1:10) and
$0.31\pm 0.08$ (1:100). This trend is in good agreement with
\citet{emqI}. Figure~\ref{fig:fsat} shows a comparison of our results
with recent estimates of the merger fraction of massive galaxies from
the literature. All merger fractions correspond to the same search
criterion in projected distance, namely $\Delta r_\perp\leq
100h_{70}^{-1}$\,kpc. In those estimates where the search radius was
smaller (typically 20-30\,kpc) , a scaling factor was included,
according to how much the fraction changes in our sample between these
two choices.  This correction is compatible with the variations
observed by \citet{LSJ:11}. Note that for major mergers ({\sl top}),
the increase at high redshift is mainly evident from the larger sample
of close pairs in COSMOS \citep{LSJ:12}, and the high redshift sample
of \citet{Bluck:12}. Our sample constrains best the evolution of the
minor merger ($\mu>0.01$) fraction, which, in combination with the
SDSS data (Ruiz et al., in preparation), suggests no redshift 
evolution in the fraction of 1:100 minor mergers out to z$\simlt$1.3.
Regarding the morphology of massive galaxies with satellites, we find
a similar distribution as with respect to the parent sample of massive
galaxies: 49\% of the massive galaxies with satellites have an
early-type morphology.  In contrast, massive galaxies with a late-type
morphology account for 39\% of the sample of systems with satellites,
and irregulars constitute the remaining 12\%.

\subsection{Border Effects}

In order to assess the effects of the edges of the SHARDS footprint,
we consider the number of sources within an aperture with radius
R=120\,arcsec centered in each of the massive galaxies selected in our
sample. This choice represents the worst case scenario: a
$500h_{70}^{-1}$\,kpc aperture extends over an angle between 60 and
112\,arcsec for the redshift range probed by our sample. We compare the
number of SHARDS-detected sources, irrespective of redshift, between
the massive galaxy under consideration, and the average obtained from
500 random positions within a region always inside the SHARDS
footprint \citep[see Fig.~2 of][]{SHARDSI}.  Only for $9$ out of the
$238$ massive galaxies ($\simlt 4$\%) a correction is needed at the
level of $\sim 8\pm 6$\%. For the fiducial $\Delta
r_\perp=100h_{70}^{-1}$\,kpc chosen to define nearby satellites, border
effects are negligible, with none of the massive galaxies requiring a
correction.

%%%%%%%%%%%%%%%%%%%%%%%%%%%%%%%%%%%%%%%%%%%%%%%%
\section{The ages of the progenitor mergers with massive galaxies}
\label{sec:ages}

The top panels of Fig.~\ref{fig:ages} ({\sl left}) show the ages of host (red
circles) and companions (blue crosses) against the stellar mass
ratio. The sample is split with respect to redshift. The hosts with a
spheroidal morphology are shown as solid circles, and disc or
irregular morphologies are shown as open circles.  Characteristic
error bars for the stellar age are given on the top-right panel for
reference. The bottom panel shows this result as an age difference
between host and companion, with the symbol notation representing
morphology as in the top panels. There is a subtle, but noticeable
difference in the ages of the host galaxies with respect to
morphology. Notice the strong trend between age difference and mass
ratio. Minor mergers -- below 1:10 -- have an age difference larger
than $\sim$1\,Gyr, except for a few pairs. This trend is consistent
with the findings of \citet{emqII}, although we note that in the
present study we have a higher constraining power on stellar ages
from the use of the SHARDS photo-spectra.

Fig~\ref{fig:Klim} showed that a population of very old and low-mass
galaxies could affect the completeness level of the sample, and thus
the conclusions drawn here. To test this issue further, we compared
the age difference between host and companion (Fig.~\ref{fig:ages},
left), with respect to the apparent magnitude of the satellite
galaxy. No significant excess of old satellites is found as we probe
fainter fluxes. For instance, taking $\Delta$Age$<1$\,Gyr to define
the fraction of systems with older satellites, we find 36\% of pairs
with older satellites over the $K_{\rm AB}=22-23.5$ range, and 16\% at
$K_{\rm AB}=23.5-25$. If a bias were present, we would have expected
an {\sl increase} of the fraction of older companions towards fainter
fluxes.  Therefore, no such bias is expected in the derivation of the
age difference down to mass ratios $\mu\sim 0.01$.

Is the trend between age difference and mass ratio of
Fig.~\ref{fig:ages} a trivial extension of the strong correlation
between mass and age? (see Fig.~\ref{fig:MvsAge}).  We explore this
issue by using the rank of the age of the satellite galaxies: for a
given pair, we compare the age of the satellite with the distribution
of stellar ages of galaxies with similar stellar mass and
redshift. The rank of the chosen satellite within this distribution
tells us whether it is significantly older/younger than similar types
of galaxies. The average rank is $59\pm28$\% without any significant
variation with respect to redshift or mass ratio, confirming that the
observed age difference is the expected one from the mass-age
correlation of Fig~\ref{fig:MvsAge}. In other words, the ages of our
satellites do not have any systematic variations caused by (or related
to) their being in close proximity to a more massive galaxy.

Fig.~\ref{fig:ages} ({\sl right}) shows the amount of rejuvenation of the
central -- if the satellite is assumed to eventually merge with it --
weighted with respect to stellar mass (top) or rest-frame $B$-band 
luminosity (bottom). The rejuvenation age is defined here as:
\begin{equation}
\Delta t_w=t_{\rm AFTER,w}-t_{\rm CEN,w}=\frac{w_{\rm SAT}}{w_{\rm SAT}+w_{\rm CEN}}
\big(t_{\rm SAT}-t_{\rm CEN}\big)
\end{equation}
where $t_{\rm CEN},t_{\rm SAT}$ are the stellar ages of central and
satellite galaxy, respectively, derived from the analysis, and $w_{\rm
  CEN},w_{\rm SAT}$ are the weights chosen. A more physically
motivated weight is the stellar mass ($w_i=M_{s,i}$, top panels of
Fig.~\ref{fig:ages}, right). We also give the luminosity-weighted
values, estimated in the rest-frame $B$-band, i.e. $w_i=L_{B,i}$
(bottom panels).  The sample is split at the median redshift.  The
figure shows that the effect of these mergers is relatively
unimportant for the merged system, with age differences below $\sim
1$\,Gyr in the majority of the close pairs, even when weighted with
respect to luminosity. This result is consistent with the overall old
stellar populations found in nearby massive galaxies \citep[see,
  e.g.,][]{Thomas:05,IGDR:11}.

\begin{figure}
  \includegraphics[width=8.5cm]{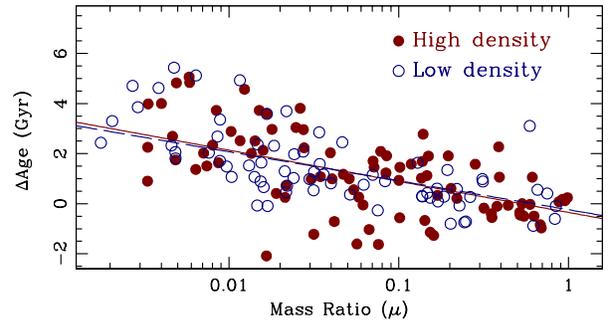}
  \caption{Effect of environment density on the age difference between
    central and satellite galaxy. Each point corresponds to a massive
    galaxy from the sample of close pairs. Density is defined for each
    one as the number of galaxies with stellar mass $M_s\geq 5\times
    10^9M_\odot$, located within a projected distance of $\Delta
    r_\perp \leq 500h_{70}^{-1}$\,kpc inside the redshift accuracy
    interval. The sample is split at the median value ($N_{500}=8$)
    into high (solid red) and low-density (open blue) regions.  For
    reference, the red solid (blue dashed) line traces the least
    squares linear fit to galaxies in high- (low-) density
    regions.}
  \label{fig:dens}
\end{figure}

Finally, we explore the impact of environment in our results. We
define $N_{500}$ as the number of galaxies with $M_s\geq 5\times
10^9M_\odot$ at a projected distance $\Delta r_\perp\leq 500h_{70}^{-1}$\,kpc of a
chosen host galaxy, within the accuracy of the photometric redshift
(at the $2\,\sigma$ level) $|\Delta z|/(1+z)\leq 0.011$. The choice of
stellar mass is justified by the completeness level of the SHARDS
sample within our redshift range for a large range of stellar
populations (see Fig.~\ref{fig:Klim}). In our
sample, the distribution suggests $N_{500}=8$ as the value that
separates between low and high density. Fig.~\ref{fig:dens} shows the
central/satellite age difference with respect to mass ratio, splitting
the sample according to galaxy density (those in high density regions
are shown as filled circles). No significant segregation is found.

%%%%%%%%%%%%%%%%%%%%%%%%%%%%%%%%%%%%%%%%%%%%%%%%
%%%%%%%%%%%%%%%%  Figure 8   %%%%%%%%%%%%%%%%%%%
%%%%%%%%%%%%%%%%%%%%%%%%%%%%%%%%%%%%%%%%%%%%%%%%
%%%%%%%%%%%%%%%%%%%%%%%%%%%%%%%%%%%%%%%%%%%%%%%%

%%%%%%%%%%%%%%%%%%%%%%%%%%%%%%%%%%%%%%%%%%%%%%%%
\section{Mass growth from \lowercase{z}$\sim$1}

\begin{figure*}
  \includegraphics[width=8.5cm]{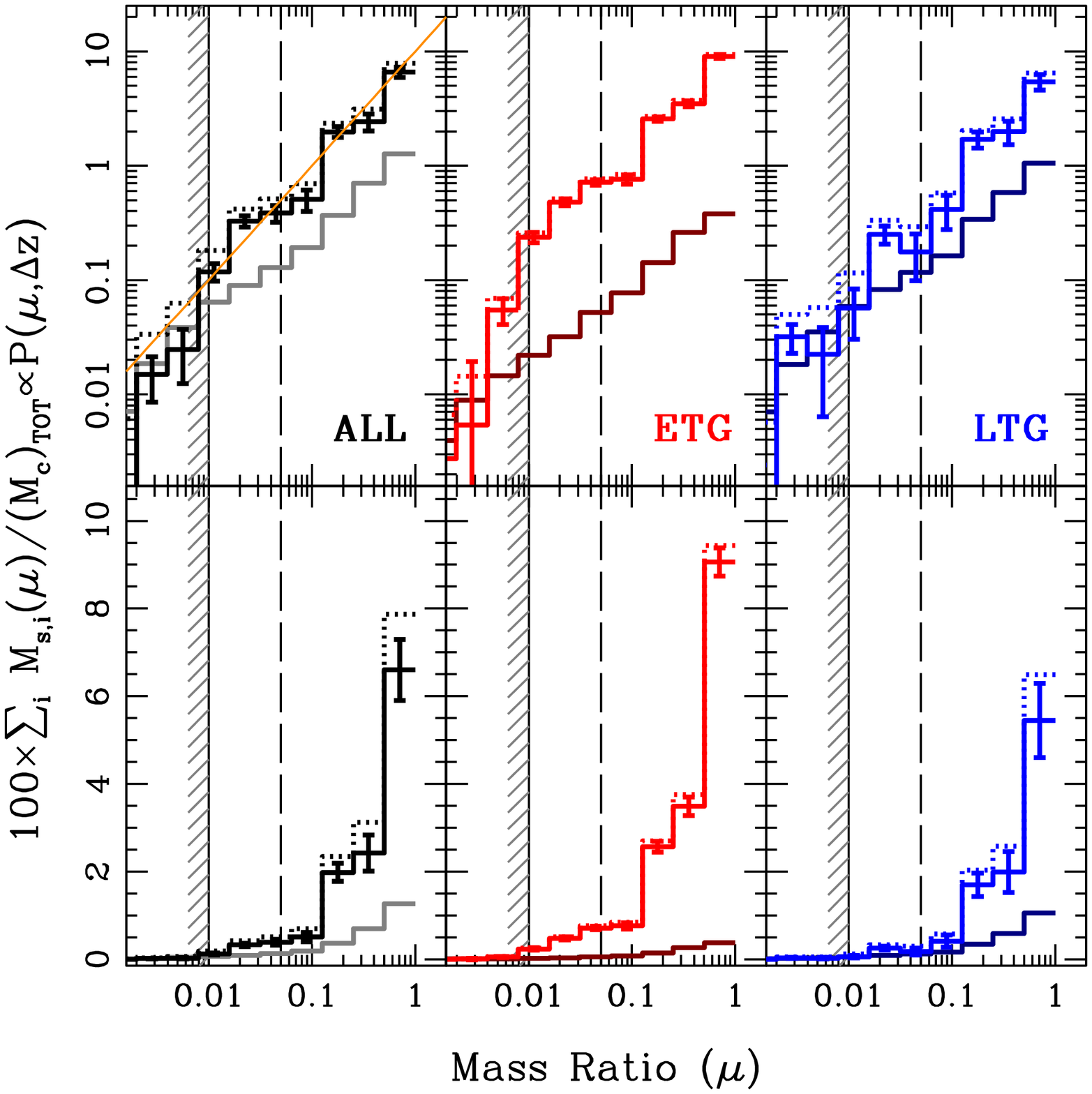}
  \includegraphics[width=8.5cm]{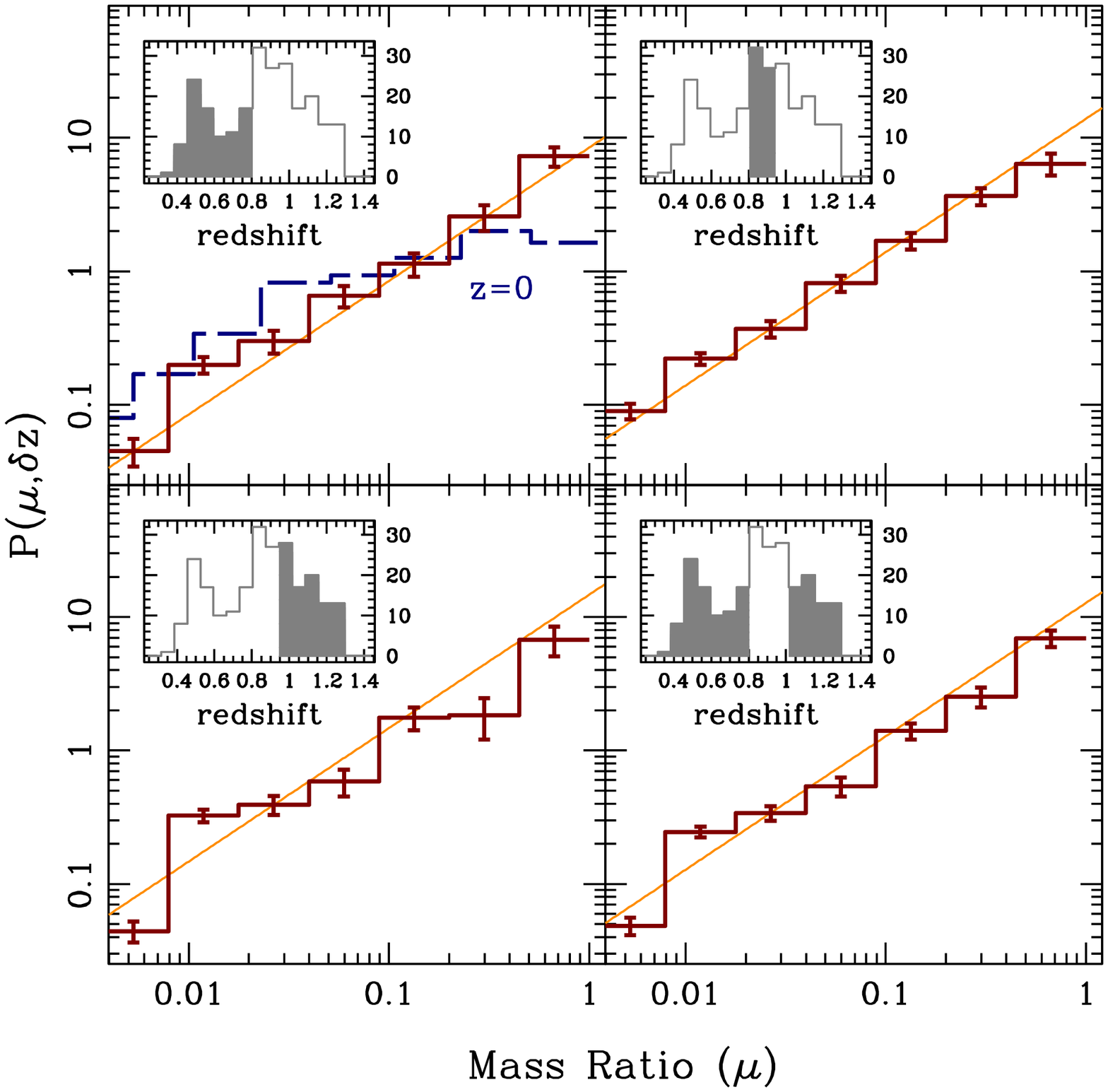}
  \caption{({\sl Left}): Histogram of the integrated mass fraction
    contributing to the growth of galaxies as a function of their mass
    ratio. Along with the complete data set ({\sl left}), The sample
    is also segregated with respect to the morphology of the central
    galaxy (early-type galaxies, {\sl middle}; and late-type galaxies,
    {\sl right}), following the same classification criterion as in
    Fig.~\ref{fig:ages}. The solid line histograms with error bars
    correspond to the final values, after applying a correction from
    the expected background contamination (histograms without error
    bars). For reference, the uncorrected results are shown as dotted
    lines. The vertical dashed line marks the conservative 1:20 limit
    of our sample, and the shaded region marks a more realistic 1:100
    limit if we consider that low-mass galaxies are dominated by young
    stellar populations. For ease of the analysis of the tail and the
    peak of the distribution, we show log-log plots ({\sl top}) and
    log-linear plots ({\sl bottom}), respectively. The orange line in
    the top-left panel is not a fit to the data. It corresponds to a
    linear increase in the mass fraction (i.e. $P(\mu)\propto\mu$).
    ({\sl Right}): The distribution is split in several redshift
    bins. The insets show as a shaded region the redshift range
    considered in each case. The sample is split in three bins at
    equal number of massive galaxies per bin. The last panel
    (bottom-right) shows the full sample, but excluding the redshift
    range $\Delta z=0.8-1.0$, where an overdensity is evident. The
    orange lines illustrate the {\sl ansatz} $P(\mu)\propto
    \mu$. In the low redshift panel, the z=0 blue dashed histogram corresponds
    to the analysis of massive galaxies from the SDSS of \citet{RTMQ:14}.}
  \label{fig:MHist}

\end{figure*}

The sample presented here can be used to quantify the dominant growth
channel of massive galaxies from z$\sim$1. Numerical simulations of
structure formation reveal an excess of satellite galaxies
\citep{Quilis:12}, with the preferred growth in mergers with a mass
ratio around 1:5 \citep{Oser:12}. Fig.~\ref{fig:MHist} ({\sl left}) quantifies
the contribution to the growth of massive galaxies as a function of
the mass ratio between the satellite and the central galaxy
($\mu$).  The histograms with error bars show the fractional
contribution of the total mass of the satellites, within each bin in
$\mu$, to the {\sl total} stellar mass in centrals. If we assume
that all satellite galaxies will be accreted into their centrals by
redshift zero, the histogram gives a probability distribution
function, P$(\mu)$, for the merger growth channel with respect to
satellite/central mass ratio, within the redshift range covered
(i.e. $\Delta z=0.3-1.3$). This interpretation is 
applicable when the mass growth rate is relatively constant over this
period, a point that is justified below. However, for major mergers
such an assumption may break down, given the evolution
with redshift claimed by some authors \citep[see, e.g.,][]{LSJ:12}.

%%%%%%%%%%%%%%%%%%%%%%%%%%%%%%%%%%%%%%%%%%%%%%%%
%%%%%%%%%%%%%%%%  Figure 9   %%%%%%%%%%%%%%%%%%%
%%%%%%%%%%%%%%%%%%%%%%%%%%%%%%%%%%%%%%%%%%%%%%%%
%%%%%%%%%%%%%%%%%%%%%%%%%%%%%%%%%%%%%%%%%%%%%%%%

From left to right, the panels of Fig.~\ref{fig:MHist} ({\sl left}) refer to
the total sample, early- and late-type central galaxies,
respectively. Poisson error bars are included.  The top panels are
log-log plots focusing on the behaviour at the low-mass end, revealing
an extended power law.  The bottom panels of Fig.~\ref{fig:MHist}
({\sl left}) plot the same histogram with linear-log axes, to show in more
detail the region around the peak of the histograms. In order to
assess the systematics regarding the selection of spurious pairs, we
ran a Monte Carlo simulation where 1000 realizations were made with
the positions of the massive galaxies being moved randomly within the
SHARDS footprint. The histograms without error bars correspond to the
average of these simulations. The uncorrected histograms, for
reference, are also shown with dotted lines. Notice that there is a
clear preference towards mass ratios $\mu\simgt 0.3$, with a
subdominant -- although measurable -- contribution from minor mergers,
that follows a linear scaling with $\log\mu$ (orange line in the
top-left panel). The simulations give a different relative
contribution with respect to $\mu$, confirming that the observed
trend is not a spurious signal from the background distribution of
sources. We should emphasize here that the histograms of
Fig.~\ref{fig:MHist} are valid on a {\sl relative} sense, i.e. with
respect to the different stellar mass ratios, with the assumption that
the observed sample provides a faithful representation of massive
galaxies over the z$\simlt$1 range.

Given that the sample only extends over a relatively small area
(130\,arcmin$^2$), we explore the effect of cosmic variance by slicing
the catalogue in three redshift bins with equal number of massive
galaxies per bin (Fig.~\ref{fig:MHist}, {\sl right}).  An additional
panel (bottom-right) also shows the effect of removing from the sample
those galaxies within the overdensity at $z\sim 0.8$. In all cases
there is a clear preponderance of major mergers, in agreement with the
trend shown in Fig.~\ref{fig:MHist} ({\sl left}).  We also estimate
the uncertainties from cosmic variance following the prescription of
\citet[][equation 27]{LSJ:14}, which depends on the volume probed, the
number density of host galaxies, and the ratio of the number density
of host (i.e. massive) galaxies and those sources from which the
companion galaxies are extracted (i.e.  the general population of
SHARDS galaxies). In this case, fluctuations in the pair fraction of
order $\pm 15$\% could be expected from cosmic variance. Such
variations are unlikely to change qualitatively our conclusions. The
orange lines in Fig.~\ref{fig:MHist} ({\sl right}) -- corresponding to
the {\sl ansatz} $P(\mu)\propto \mu$ -- are fully consistent with our
results. For reference, the recent work of \citet{RTMQ:14} based on a
large sample of z$\sim$0 massive {\sl elliptical} galaxies from the Sloan
Digital Sky Survey is shown in the top-left panel as a dashed blue
histogram. Note the significant deficit at z$\sim$0 in the contribution from
major mergers. Since the SDSS sample only includes elliptical galaxies,
this evolution in the contribution from major mergers is expected to be
even more significant for the general population of massive galaxies
(Ruiz et al., in preparation).

\begin{figure}
  \includegraphics[width=8.5cm]{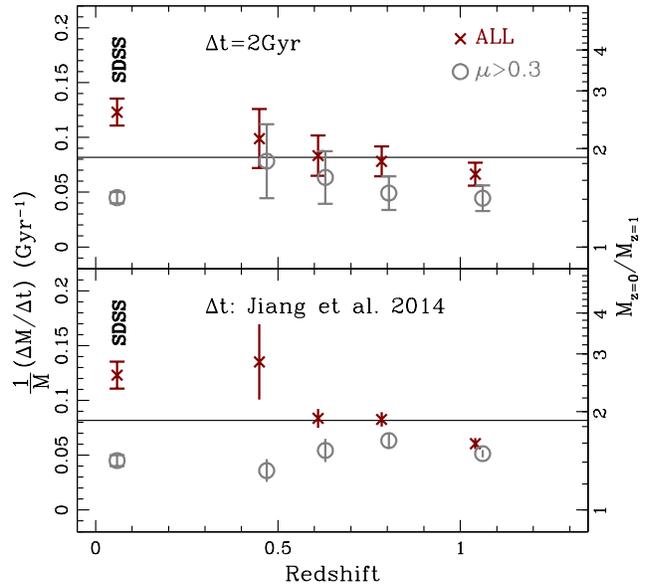}
  \caption{{\sl Top:} Growth rate of massive galaxies with respect
    to redshift for the simple case of a fixed ($\Delta t=2$\,Gyr)
    merging timescale. The red crosses (grey circles) correspond to
    the contribution from the whole sample (only from major mergers,
    $\mu>0.3$).  The axis on the right gives the expected stellar mass
    growth between z=1 and z=0 assuming a constant rate. For
    reference, the points on the left, labelled SDSS, correspond to a
    nearby sample ($z\sim 0.06$) using the Sloan Digital Sky Survey
    \citep{RTMQ:14}. {\sl Bottom:} Same as above, although taking into
    account the dependence of the merging timescale on projected 
    separation and mass ratio, according to the prescription of
    \citet{Jiang:14}.}
  \label{fig:MGrowth}
\end{figure}

We derive a simple estimate for the mass growth of massive galaxies
over the redshifts probed, by calculating the ratio between the
stellar mass in satellites and centrals, within an interval $\Delta
t=2$\,Gyr, roughly corresponding to an average merger timescale
\citep[see, e.g.,][]{KW:08}. This is meant to be a rough estimate that
assumes a constant inflow of satellites with time, as supported by the
mild evolution in the fraction of close pairs out to z=1.3
(Tab.~\ref{tab:sims}). Fig.~\ref{fig:MGrowth} (top) shows the result
for the complete sample (red crosses) and the subsample involving
only major mergers ($\mu>0.3$, grey open circles). The measurements
include Poisson error bars. Within uncertainties, the mass growth does
not change significantly with redshift, with a value around
$\tau^{-1}\equiv (\Delta M/M)/\Delta t\sim 0.08\pm 0.02$\,Gyr$^{-1}$.
We note that when segregating the sample based on our visual
morphological classification of early- and late-type galaxies, no
significant difference is found in their respective mass growth rate.
Such a rate would imply a cumulative increase in the stellar mass
between z=1 and z=0 of
\begin{equation}
\left(\frac{M_{\rm z=0}}{M_{\rm z=1}}\right)=
e^{(t_{\rm z=0}-t_{\rm z=1})/\tau}\sim 1.9\pm 0.3
\end{equation}
For reference, the estimate of \citet{VdK:10},
$\Delta\log M/\Delta z  =-0.15$,
extended out to z=1 gives a mass growth in massive
galaxies of a factor $\approx 1.4$. We note this result is based on a
different approach, namely a comparison of the mass budget at low and
high redshift. Our results are also consistent with the
$M_{z=0}/M_{z=1}=1.3$ mass growth factor in the COSMOS field
\citep{LSJ:12}, if we take into account that this estimate extends
only to $\mu>0.1$ mergers, selecting close pairs out to $\Delta
r_\perp<30h^{-1}$\,kpc.

In the following, we consider a more realistic estimate of mass
growth, by taking into account the merger timescales of close pair
systems, which mainly depend on the pair separation and the mass
ratio. In \citet{KW:08} simulated close pairs from the Millennium
database were used to derive an estimate of merging timescales, given
as a function of pair separation, stellar mass of the central galaxy
and redshift. More recently, \citet{Jiang:14} revisited these
estimates and proposed an expression that uses the virial masses --
given that dynamical friction should be dependent on these -- and
extend their analysis to lower mass ratios, with an expression where
the merging timescale depends linearly on this ratio.
Fig.~\ref{fig:MGrowth} ({\sl bottom}) assumes a merger timescale
following the Eq.~7 in \citet{Jiang:14}, using the parameterisation of
\citet{Moster:10} to convert stellar masses into virial masses, as a
function of redshift. Note that no large differences are apparent,
within error bars, with respect to the assumption of instantaneous
merging (Fig.~\ref{fig:MGrowth}, {\sl top}), because most of the
contribution to the mass growth is in major mergers, for which the
merging timescales are short. There is a noticeable trend of a
diminishing role of major mergers at lower redshift.  The relative
contribution from major mergers ($\mu>0.3$) is
quite significant, amounting to $71\pm 8$\% of the total growth probed
by this sample. The analysis of \citet{Bluck:12} of the progenitors of
mergers involving massive galaxies at higher redshifts (z=1.7--3) lead
to a factor $3$ in stellar mass growth, with 65\% coming from major
mergers. \citet{LSJ:12} find at $z<1$ a 75\% fraction from
$\mu>0.25$. Our results are fully consistent \citep[see also ][]{Tal:13}.

%%%%%%%%%%%%%%%%%%%%%%%%%%%%%%%%%%%%%%%%%%%%%%%%
\section{Conclusions}

Satellite galaxies in the region of influence of massive galaxies over
a wide range of redshifts provide valuable information about the open
question regarding the size growth over the past 8-10\,Gyr of cosmic
history \citep[e.g.][]{Truj06,Truj07}. In this paper we build a sample
of massive galaxies from the deep SHARDS dataset.  Photometric data
from medium band filters allow us both to obtain enough accuracy in
the photometric redshift estimates to minimise contamination ($|\Delta
z|/(1+z)\sim 0.55$\%, median, confirmed with a spectroscopic sample
down to $K_{\rm AB}\sim 24$) and to constrain the age of the stellar
populations by use of a large volume of synthetic models. The SHARDS
sample is robustly complete over the 0.3$<$z$<$1.3 redshift range down
to 1:20 if all stellar ages are allowed for, or down to 1:100 if we
consider that galaxies with the lowest masses will be significantly
younger (Fig.~\ref{fig:Klim}).

%%%%%%%%%%%%%%%%%%%%%%%%%%%%%%%%%%%%%%%%%%%%%%%%
%%%%%%%%%%%%%%%%  Figure 10  %%%%%%%%%%%%%%%%%%
%%%%%%%%%%%%%%%%%%%%%%%%%%%%%%%%%%%%%%%%%%%%%%%%
%%%%%%%%%%%%%%%%%%%%%%%%%%%%%%%%%%%%%%%%%%%%%%%%

From the final sample of $238$ massive galaxies (M$_s\geq
10^{11}$\,M$_\odot$), we select those with a nearby companion -- at a
projected distance $\Delta r_\perp\leq 100h_{70}^{-1}$\,kpc, and at
the same redshift within the (2\,$\sigma$) accuracy of the
estimates. The fraction of close pairs is consistent with the results
in the literature. However, SHARDS is the first survey capable of
probing the evolution of the minor merger rate $\mu>0.01$ out to
$z\sim 1$, with high completeness. In conjunction with the local
estimate of the minor merger fraction from the Sloan Digital Sky
Survey (Ruiz et al., in preparation), we find no significant evolution
in the fraction of 1:100 minor merger progenitors out to z$\simlt 1.3$
(Fig.~\ref{fig:fsat}).

The ages of the stellar populations show a strong correlation
with mass fraction between host and companion, with a significant age difference for
the more unequal mass mergers (Fig.~\ref{fig:ages}). However, this
trend is expected from the general mass-age correlation
(Fig.~\ref{fig:MvsAge}), i.e.  there is no significant difference
between the ages of the companions and the ages of the general
population at fixed stellar mass.  No significant segregation is
observed either when splitting the sample with respect to local
density, defined as the number of massive galaxies within a projected
distance of $500h_{70}^{-1}$\,kpc, and at the same redshift
(Fig~\ref{fig:dens}). We emphasize that since massive galaxies are
targeted in this study, our sample probes halos where environmental
effects should be maximal at those redshifts. Comparing with the
results from groups at lower redshift \citep[see,
  e.g.,][]{vdBosch:08}, our data suggest that quenching in low-mass
satellites is not a highly efficient mechanism, and may take
significantly longer times, comparable with the lookback times of the
systems under study (4--8\,Gyr).

The contribution from satellites to the growth of massive galaxies is
found to originate mainly from a mass ratio $\mu\simgt 0.3$
(Fig.~\ref{fig:MHist}), with a net increase to the stellar mass
between z=1 and 0 of a factor $\sim 1.9$, or an average mass growth rate
$(\Delta M/M)/\Delta t\sim 0.08\pm 0.02$\,Gyr$^{-1}$, which should be
taken as an upper limit to the growth via mergers, as this paper
assumes that all central-satellite systems will end up as a merging
event. A fraction $\sim 71$\% of this mass growth is in the form of
mergers with a mass ratio $\mu>0.3$.  The combination of age and
mass difference in the merger progenitors studied here would imply a
rejuvenation of massive galaxies $\simlt 1$\,Gyr over the redshift
range probed (Fig.~\ref{fig:ages}). Hence, we find that the dominant
channel of mass growth does not introduce any large variations in the
age distribution of the stellar material incorporated at z$\simlt
1$. This result is in agreement with the small radial age gradients
observed in massive early-type galaxies at moderate redshift
\citep{ferr09} and at z$\simlt$0.1 \citep{FLB:11,FLB:12}. The small
amount of rejuvenation is also consistent with the lack of
an age segregation on the mass-size plane \citep{I3}.

The observations of merger progenitors over the past 8\,Gyr of cosmic
time impose strong constraints on models of galaxy formation and
evolution. The current ``paradigm'' seems to converge towards a
significant contribution from minor mergers in massive galaxies at
z$\simlt$1 \citep[see, e.g.][]{Oser:12,Lackner:12,Shankar:13} --
mainly finding an explanation to the evolution on the mass-size plane
\citep[see, e.g.][]{Naab:09}. However, the observations presented here and
elsewhere \citep[see, e.g.][]{LSJ:11,Bluck:12} suggest a substantial
revision of the recipes introduced to describe the ''baryon physics''
in numerical simulations.

%%%%%%%%%%%%%%%%%%%%%%%%%%%%%%%%%%%%%%%%%%%%%%%%
\section*{Acknowledgments}
We thank Carlos L\'opez-Sanjuan for his very valuable comments and
suggestions.  SHARDS is currently funded by the Spanish MICINN/MINECO
with grant AYA2012-31277. Based on observations made with the Gran
Telescopio Canarias (GTC), installed at the Spanish Observatorio del
Roque de los Muchachos of the Instituto de Astrof\'\i sica de
Canarias (IAC), in the island of La Palma. The authors acknowledge the use
of the UCL Legion High Performance Computing Facility (Legion@UCL),
and associated support services, in the completion of this work. IF
acknowledges support from the IAC to carry out this research project.
EMQ acknowledges the support of the European Research Council
via the award of a Consolidator Grant (PI McLure).
AH-C acknowledges support from the Augusto Gonz\'alez Linares
programme (Universidad de Cantabria), and the Spanish ``Plan
Nacional'' grant AYA2012-31447.  This work has been supported by the
`Programa Nacional de Astronom\'\i a y Astrof\'\i sica' of the Spanish
Ministry of Science and Innovation under grant
AYA2010-21322-C03-02. We have made extensive use of the Rainbow Cosmological
Surveys Database, operated by the Universidad Complutense de Madrid (UCM),
in partnership with the University of California Observatories at Santa
Cruz (UCO/Lick, UCSC).

%%%%%%%%%%%%%%%%%%%%%%%%%%%%%%%%%%%%%%%%%%%%%%%%
%%%%%%%%%%%%%%%   REFERENCES   %%%%%%%%%%%%%%%%%%%%%%
%%%%%%%%%%%%%%%%%%%%%%%%%%%%%%%%%%%%%%%%%%%%%%%%

%%%%%%%%%%%%%%%%%%%%%%%%%%%%%%%%%%%%%%%%%%%%%%%%%%%%%%%%%%%
\appendix

\setcounter{figure}{0}
\renewcommand{\thefigure}{A\arabic{figure}}

%%%%%%%%%%%%%%%%%%%%%%%%%%%%%%%%%%%%%%%%%%%%%%%%
%%%%%%%%%%%%%%%%  Figure A1   %%%%%%%%%%%%%%%%%%%
%%%%%%%%%%%%%%%%%%%%%%%%%%%%%%%%%%%%%%%%%%%%%%%%
\begin{figure*}
  \includegraphics[width=80mm]{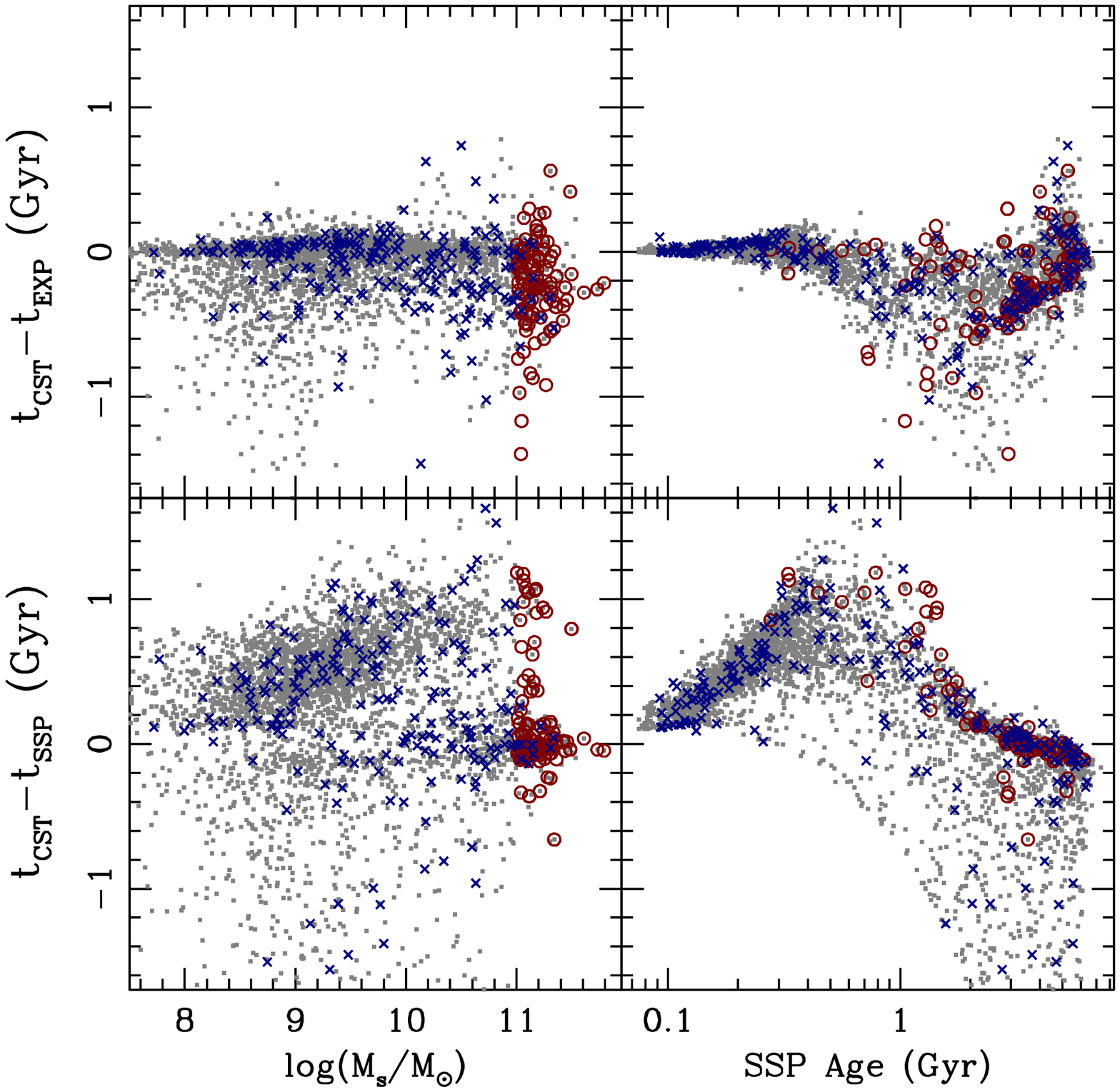}
  \includegraphics[width=80mm]{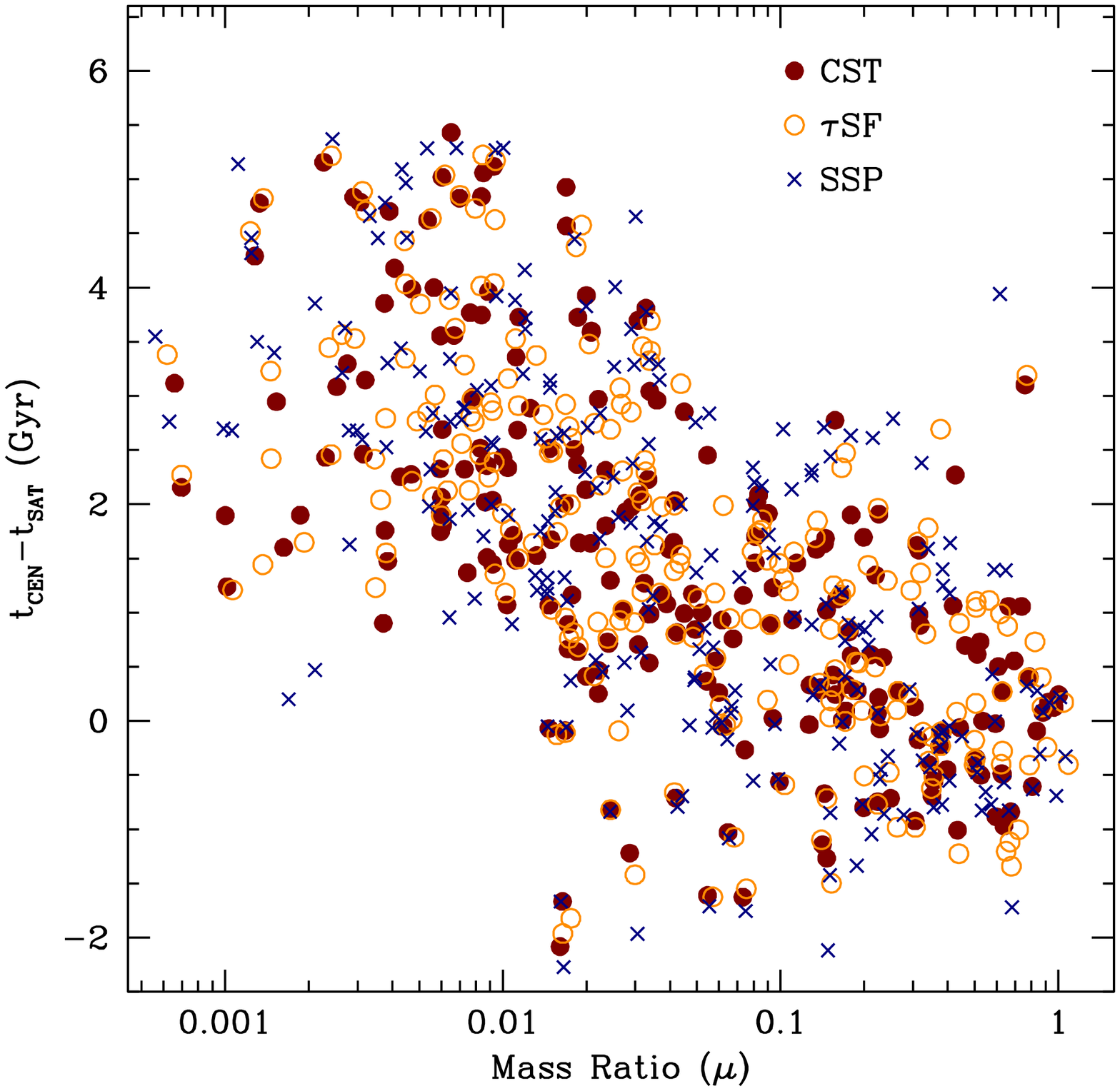}
  \caption{{\sl Left}: Comparison of stellar ages between the three different
    model grids used in this paper (Tab.~\ref{tab:pars}). The grey dots
    correspond to the general sample, whereas host (companion) galaxies
    in close pairs are shown as red open circles (blue crosses).
    {\sl Right:} Comparison of the correlation between mass ratio and age difference
    for the three model grids, as labelled.
  }
  \label{fig:comp}
\end{figure*}
%%%%%%%%%%%%%%%%%%%%%%%%%%%%%%%%%%%%%%%%%%%%%%%%

\section{Effect of the star formation history on derived parameters}

Tab.~\ref{tab:pars} shows that three different sets of 
(standard) grids were chosen to parameterise the star formation histories
of galaxies. We show here the differences found between them with
respect to stellar age and mass. All three models are explored in the same
way, with the same base population synthesis models, i.e.
\citet{BC03} models with a \citet{Chab:03} initial mass function.

Fig.~\ref{fig:comp} ({\sl left}) plots the difference between the best
fit stellar ages of the three different as a function of stellar mass
({\sl left}) or age ({\sl right}). Note the expected trend towards
younger SSP-ages and older $\tau$-model ages. The latter is mainly
caused by the tendency of the fitting model to favour small timescales
in order to avoid the extended tail of the exponential function
\citep[see][for a detailed example corresponding to a massive galaxy
  at high redshift]{FW4871}. The grey dots correspond to the general
population, whereas the sample of galaxies in close pairs are
highlighted as red open circles (host) and blue crosses (lower mass
companion). The figure shows that the sample of galaxies in close
pairs explored in this paper is representative of the general
population, so no systematic bias is expected with respect to the
choice of star formation history.

In addition, we present on the right hand side of Fig.~\ref{fig:comp}
the trend between mass ratio and age difference, i.e. the equivalent of
Fig.~\ref{fig:ages} (bottom-left panels) with all three model grids,
to confirm that the strong correlation is independent of the 
parameterisation of the star formation history.

\label{lastpage}
\end{document}